\documentclass[prl,aps,superscriptaddress,footinbib,twocolumn]{revtex4-2}
\usepackage{times}
\usepackage[colorlinks=true,urlcolor=blue,citecolor=blue]{hyperref}
\usepackage{soul}
\usepackage{graphicx}
\usepackage{amsmath}
\usepackage{mdframed}
\usepackage{rotating}
\newcommand{\figpath}{./}

\begin{document}
	
    \title{Quantum automated theorem proving}
	
    \author{Zheng-Zhi Sun}
    \affiliation{Center for Quantum Information, IIIS, Tsinghua University, Beijing 100084, China}
    \author{Qi Ye}
    \affiliation{Center for Quantum Information, IIIS, Tsinghua University, Beijing 100084, China}
      \affiliation{Shanghai Qi Zhi Institute, Shanghai 200232, China}
    \author{Dong-Ling Deng}
    \email{dldeng@tsinghua.edu.cn}
    \affiliation{Center for Quantum Information, IIIS, Tsinghua University, Beijing 100084, China}
      \affiliation{Shanghai Qi Zhi Institute, Shanghai 200232, China}
    \affiliation{Hefei National Laboratory, Hefei 230088, China}

    \begin{abstract}
    Automated theorem proving, or more broadly automated reasoning, aims at using computer programs to automatically prove or disprove mathematical theorems and logical statements. It takes on an essential role across a vast array of applications and the quest for enhanced theorem-proving capabilities remains a prominent pursuit in artificial intelligence. Here, we propose a generic framework for quantum automated theorem proving, where the intrinsic quantum superposition and entanglement features would lead to potential advantages. In particular, we introduce quantum representations of knowledge bases and propose corresponding reasoning algorithms for a variety of tasks. We show how automated reasoning can be achieved with quantum resolution in both propositional and first-order logic with quadratically reduced query complexity. In addition, we propose the quantum algebraic proving method for geometric theorems, extending Wu’s algebraic approach beyond the classical setting. Through concrete examples, including geometry problems from the International Mathematical Olympiad, we demonstrate how a quantum computer may prove geometric theorems with quadratic better query complexity. Our results establish a primary approach towards building quantum automatic theorem provers, which would be crucial for practical applications of both near-term and future quantum technologies. 
    \end{abstract}

    \maketitle

\noindent Automated theorem proving (ATP) lies at the heart of symbolic artificial intelligence, formal verification, and mathematical reasoning \cite{Fitting2012First,Loveland2016Automated}. Its goal is to mechanize logical inference: given a set of axioms or hypotheses, an ATP system determines—through sound and complete inference rules—whether a target conclusion follows. Over the past decades, ATP has become indispensable in domains ranging from hardware and software verification \cite{Hasan2015Formal} to the formalization of modern mathematics, where machine-checked proofs now reach a level of rigor unattainable by human reasoning alone (e.g., the formalization of the four-color theorem and the Kepler conjecture)~\cite{Gonthier2008Formal,Hales2017Formal}. Recent successes include the large-scale formal verification of sophisticated mathematical results \cite{Gowers2025Conjecture} and the emergence of geometry solvers capable of handling Olympiad-level problems without human-provided demonstrations \cite{Trinh202401Solving,Guo2025Deepseek,Hubert2025Olympiad}.

Despite these advances, ATP continues to face a fundamental scalability barrier. Logical inference—whether based on resolution, unification, or algebraic elimination—typically requires exploring a vast combinatorial search space \cite{Fitting2012First,Loveland2016Automated}. As the size of the knowledge base grows, the number of candidate inference steps increases rapidly, leading to severe time and memory costs. Even state-of-the-art systems that combine symbolic deduction with learned heuristics, such as neural-guided theorem provers and recent geometry engines, remain constrained by this underlying combinatorial explosion \cite{Trinh202401Solving,Guo2025Deepseek,Hubert2025Olympiad}. In practice, much of the computational effort in ATP is spent repeatedly searching for valid inference steps, filtering out invalid candidates, and performing algebraic manipulations whose cost grows polynomially or exponentially with problem size.

      \begin{figure*}[htb]
		\includegraphics[width=1\linewidth]{\figpath/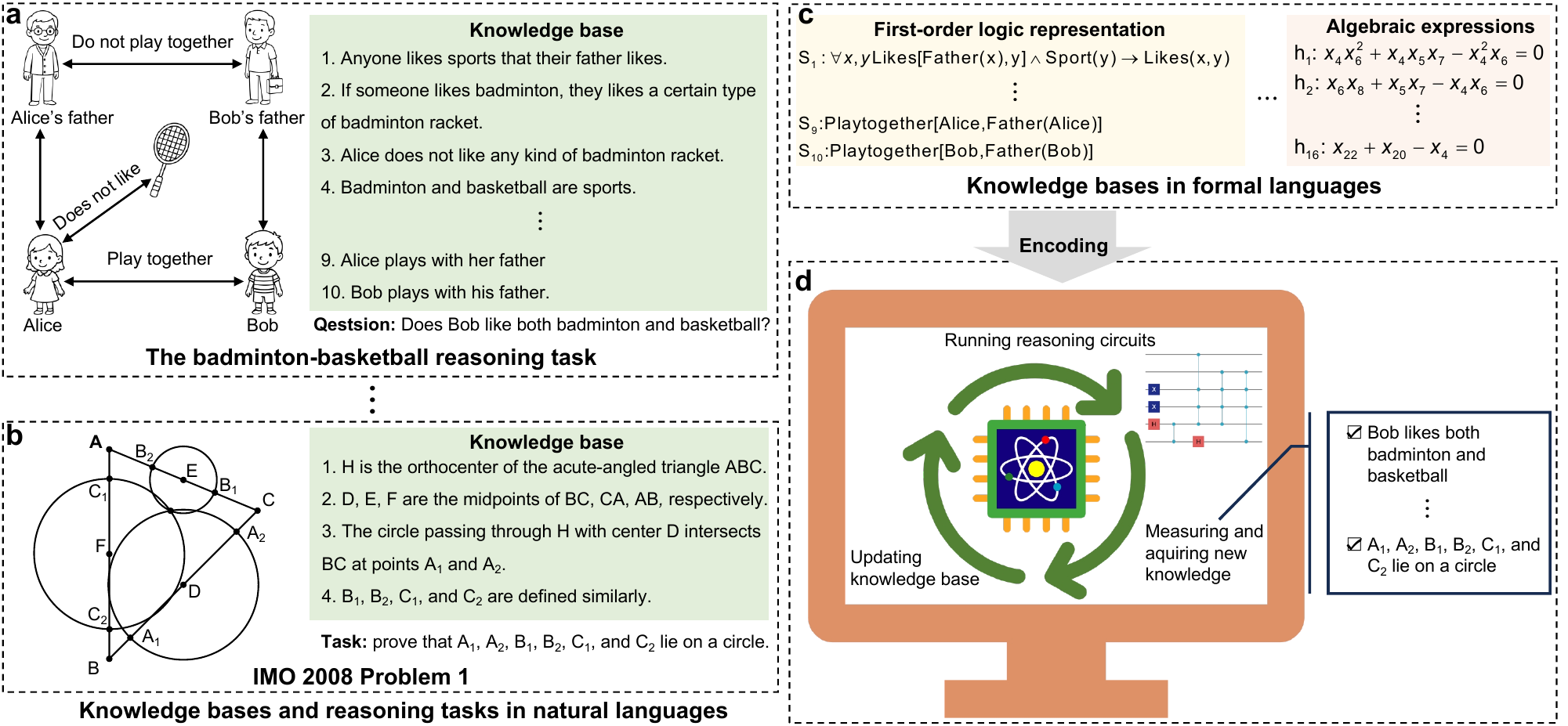}
		\caption{\label{fig-qatp-sketch} \textbf{An illustration of quantum automated theorem proving}. \textbf{a}, The badminton-basketball reasoning task expressed in natural language. The initial knowledge base in this case consists of ten sentences and the target is to infer whether Bob likes both badminton and basketball or not (see Supplementary Information Section II.B for details). \textbf{b}, The IMO 2008 Geometry Problem 1 diagram and statements. \textbf{c}, Left panel: first-order logic representation of the knowledge base for the badminton-basketball reasoning task; Right panel: algebraic expressions of the knowledge base for the IMO 2008 Geometry Problem 1. By using different encoding strategies, these knowledge bases expressed in formal languages are mapped into either quantum states or quantum circuits, depending on the specific task. \textbf{d},  Reasoning on a quantum computer.
        }
	\end{figure*}

Quantum computing \cite{Nielsen201206Quantum}, grounded in the principles of quantum mechanics, would offer a fundamentally different computational paradigm for addressing these challenges. By exploiting quantum superposition and interference, quantum algorithms can evaluate many candidate solutions simultaneously and amplify the probability of desired outcomes. Landmark developments in quantum hardware—including programmable superconducting circuits \cite{Arute201910Quantum, Wu202110Strong, Krinner202205Realizing, Kim202306Evidence,Xu202407Non,Jin2025Topological}, trapped ions \cite{Egan202110Fault, Moses202312Race}, photonic systems \cite{Zhong202012Quantum, Madsen202206Quantum}, and Rydberg-atom arrays \cite{Ebadi202107Quantum,Bluvstein2024Logical}—together with latest breakthroughs in quantum supremacy \cite{Arute201910Quantum,Zhong202012Quantum,Wu202110Strong,Madsen202206Quantum} and quantum error correction \cite{Acharya202302Suppressing,Sivak202303Real,Ni202303Beating,Bluvstein2024Logical,Gupta2024Encoding,Acharya2025, Wang2025Demonstration,Lacroix2025Scaling}, are bringing quantum computation closer to the regime where nontrivial and noise-resilient algorithms can be executed reliably. Indeed, the interplay between quantum computing and machine learning has given rise to a new research frontier of quantum machine learning \cite{Biamonte2017Quantum,Dunjko2018Machine,DasSarma2019Machine,Cerezo2022Challenges,Li2025Pitfalls}. A variety of quantum learning algorithms with potential advantages have been proposed \cite{Harrow2009Quantum,Lloyd2014Quantum,Dunjko2016Quantum,Gao2018Quantum,Liu2021Rigorous,Jerbi2023Quantum} and some of them have been demonstrated in proof-of-principle experiments \cite{Havlicek2019Supervised,Hu2019Quantum,Huang2022Quantum,Ren2022Experimental}. 
These advances motivate the question of whether quantum resources can be harnessed not only for numerical computation or optimization, but also for symbolic reasoning.

Here, we introduce quantum automated theorem proving (QATP): a general framework that integrates quantum algorithms with the core inference mechanisms of ATP while preserving the semantics of classical logic. Rather than replacing symbolic reasoning with heuristic optimization, QATP targets the dominant computational bottlenecks of theorem proving—namely, large-scale search, clause selection, and algebraic verification—by embedding them into quantum subroutines with provable performance guarantees. We develop quantum resolution algorithms for propositional and first-order logic that reduce the query complexity of inference, and we construct a quantum implementation of Wu’s method \cite{Wu1978Decision} for geometric theorem proving by reformulating polynomial manipulation and identity testing in a quantum-compatible form. Through concrete examples, including geometry problems from the International Mathematical Olympiad (IMO) \cite{IMOOfficialWebsite}, we demonstrate how a quantum computer may achieve ATP with quadratic improved query complexity.
By connecting quantum algorithms with formal logic and algebraic proof systems, this work opens a pathway toward scalable, verifiable reasoning on quantum computers and establishes a bridge between quantum computation and symbolic artificial intelligence.

\begin{figure*}[htbp]
    \centering
    \includegraphics[width=1\linewidth]{\figpath/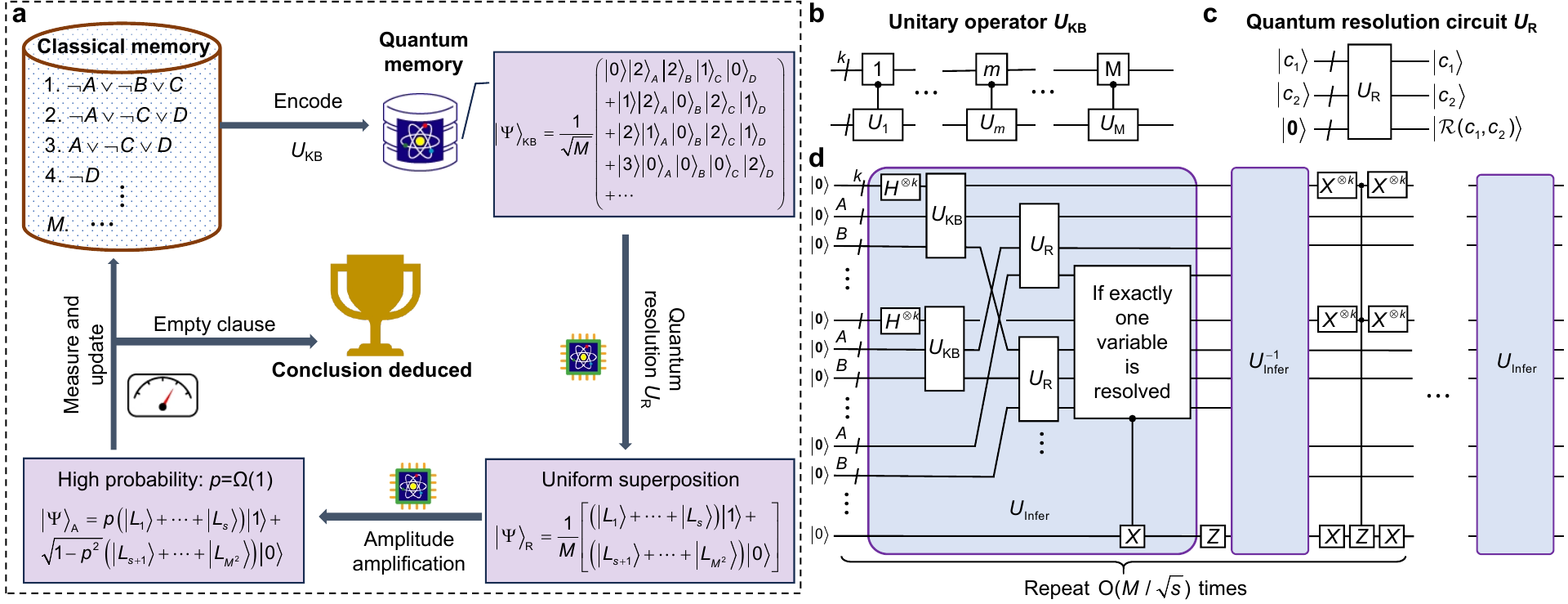}
    \caption{\label{fig-qatp-propositional-logic} 
    \textbf{Quantum resolution for propositional logic.}
    \textbf{a}, Schematic overview of the quantum resolution procedure. 
    For a given reasoning task, the knowledge base is expressed in propositional logic as a single sentence in conjunctive normal form, which is a conjunction of $M$ clauses. 
    The quantum resolution process consists of four main steps. 
    First, an appropriate encoding unitary $U_{\text{KB}}$ is constructed to load the classical knowledge base into quantum memory. 
    Applying $U_{\text{KB}}$ to the initial state $|\mathbf{0}\rangle = |0 \cdots 0\rangle$ prepares a many-body quantum state $|\Psi\rangle_{\text{KB}}$ that encodes the entire knowledge base. 
    Second, we apply the quantum resolution unitary $U_\text{R}$ to calculate the resolution results for every propositional variable, and a verification circuit to identify valid resolvents with exactly one resolved variable (see Supplementary Information Section II.A), yielding the state $|\Psi\rangle_\text{R}$. This state is a uniform superposition of $M^2$ components—$s$ of which correspond to valid clauses (state $\left| 1 \right\rangle $ on the last qubit) and the remaining $M^2 - s$ to invalid ones (state $\left| 0 \right\rangle $ on the last qubit). 
    Each clause after resolution is denoted as $L_j$ ($j = 1, \ldots, M^2$). 
    Third, amplitude amplification increases the probability of observing valid components from $s/M^2$ to $\Omega(1)$. 
    Finally, measuring the post-amplification state $|\Psi\rangle_\text{A}$ in the computational basis yields the $s$ valid clauses. 
    If an empty clause is obtained, the reasoning process terminates and the target conclusion is derived. 
    Otherwise, the newly generated clauses are added to the knowledge base, and the procedure is repeated until either an empty clause arises (the theorem is proved) or no new clause can be obtained (the theorem cannot be inferred). 
    \textbf{b}, Quantum circuit implementing the encoding unitary $U_{\text{KB}}$. This circuit consists of $M$ controlled subcircuits. Each subcircuit maps the state from $\left| \bf{0} \right\rangle $ on $N$ ququarts (bottom wire) to the state representing the $m$-th clause, conditioned on the state $\left| m \right\rangle $ on the $k=\log _2 M$ index qubits (top wire).
    \textbf{c}, Circuit representation of the quantum resolution unitary $U_\text{R}$. This circuit resolves two variables $c_1$ and $c_2$ and outputs the resolution result $\mathcal{R}({c_1}, {c_2})$ on the third ququart.
    \textbf{d}, Quantum circuit for amplitude amplification. 
    Here, $H$ denotes the Hadamard gate, and $X$ and $Z$ represent the Pauli-$X$ and Pauli-$Z$ operators, respectively. The key idea is to entangle states corresponding to valid (invalid) clauses with state $\left| 1 \right\rangle $ ($\left| 0 \right\rangle $) of the last qubit,
 and to amplify the amplitude of valid clauses. After $O(M/\sqrt{s})$ iterations,  the probability of obtaining valid clauses is increased  from $s/M^2$ to $\Omega (1)$.
    }
\end{figure*}

\vspace{.5cm}
\noindent\textbf{\large{}Notations and the general framework}{\large\par}

\noindent We first introduce some basic notations and the general framework for QATP (Fig.~\ref{fig-qatp-sketch}). Classically, ATP establishes mathematical truths through a framework of formal logic \cite{Loveland2016Automated}. This foundation comprises the syntax and semantics of logical languages, which together construct expressions that can be rigorously evaluated. These expressions populate a knowledge base, which is a repository of sentences in a formal language that enables precise inference. Key representation languages include propositional logic, which handles binary statements, and first-order logic, which extends this with quantifiers and predicates to model complex relationships. The core ATP process involves translating a conjecture into this formal framework and then systematically exploring its logical consequences by applying inference rules until the theorem is proven or refuted.

In Fig.~\ref{fig-qatp-sketch}, we sketch the general pipeline for QATP. Suppose that we have a reasoning task, such as the badminton-basketball reasoning task (Fig.~\ref{fig-qatp-sketch}\textbf{a}) or the IMO 2008 Geometry Problem 1 (Fig.~\ref{fig-qatp-sketch}\textbf{b}), which is typically described in natural language. We first represent the knowledge base of the task in a formal language (Fig.~\ref{fig-qatp-sketch}\textbf{c}). Using an appropriate encoding strategy, we map the knowledge base into a quantum state and prepare this state on a quantum computer. We then execute the reasoning quantum circuits, such as quantum resolution, on the prepared state. We measure the output states to deduce new valid sentences. If the obtained new sentences include the targeted conclusion, we terminate the process, and the conclusion is proved. Otherwise, we update the knowledge base with these new sentences and repeat the reasoning process until the conclusion is deduced or no new sentences can be obtained (Fig.\ref{fig-qatp-sketch}\textbf{d}).
In the latter case, we conclude that the target conclusion cannot be deduced from the knowledge base if the inference algorithm used is complete. In the following, we give two concrete examples of quantum reasoning, one on quantum resolution in propositional and first-order logic, and the other that extends Wu's method for proving geometric theorems into the quantum domain, to demonstrate how QATP would work and showcase potential advantages compared to their classical counterparts.

\vspace{.5cm}
\noindent\textbf{\large{}Quantum resolution}{\large\par}    

\noindent In classical ATP, one of the most fundamental and powerful inference rules is resolution \cite{Robinson196501Machine,Russell2020Artificial}. 
This rule is sound in both propositional and first-order logic, complete in propositional logic, and refutation-complete in first-order logic. It plays a central role in a wide range of applications.  In this section, we introduce quantum resolution algorithms and explore their potential quantum advantages. To begin with, we first outline the basic principles and procedures underlying the use of resolution algorithms in ATP. The resolution method is typically used in a refutation-based manner \cite{Robinson196501Machine,Russell2020Artificial}: it seeks to derive a contradiction from the axioms together with the negation of the theorem to be proved, thus indirectly establishing the validity of the theorem. This process starts by transforming the logical statements into conjunctive normal form (CNF), which is a canonical format expressing formulas as conjunctions of disjunctions of literals. 
Once in CNF, the algorithm iteratively applies the resolution rule, selecting pairs of clauses and resolving their complementary literals to generate new clauses. The basic form of resolution for clauses in propositional logic is given by
\begin{align}\label{eq-resolution-propositional}
    	{{{a_1} \vee  \cdots  \vee {a_n} \vee c,{b_1} \vee  \cdots  \vee {b_m} \vee \neg c} \over {{a_1} \vee  \cdots  \vee {a_n} \vee {b_1} \vee  \cdots  \vee {b_m}}},
    \end{align}
where $a_i$, $b_i$, and $c$ are literals, and the horizontal line denotes logical entailment. A literal is either a propositional variable or its negation—each representing a statement that can be true or false. The symbols $\vee$, $ \wedge $, and $\neg$ are logical operators known as ``disjunction'', ``conjunction'', and ``negation'', respectively.  
The resolution rule operates by eliminating complementary literals—that is, instances of the same propositional variable appearing with opposite polarities—and retaining the remaining literals from the two input clauses. A resolution step is valid only when exactly one such complementary pair is resolved. If the resulting clause is valid and not already present in the knowledge base, it is added to the knowledge base. This process continues iteratively until an empty clause representing the contradiction is derived, thus proving the original theorem. If no further resolutions are possible, the theorem cannot be inferred from the given knowledge base. Propositional resolution is a sound and complete inference method, making it a cornerstone of the classical automated theorem proving. It also serves as a canonical example for developing the quantum inference system introduced below.

To introduce QATP within the framework of propositional logic (Fig.~\ref{fig-qatp-propositional-logic}), we first represent the knowledge base as a many-body quantum state composed of four-level quantum systems (ququarts). Each ququart can be equivalently mapped onto a pair of standard qubits. For a clause involving $N$ propositional variables, we assign $N$ ququarts, where the state of each ququart encodes the status of its corresponding propositional variable—namely, absent, positive, negative, or resolved. The index of each ququart corresponds directly to that of the associated propositional variable. Since literals within a clause in CNF are implicitly connected by disjunctions, the disjunction symbols are omitted (Fig.~\ref{fig-qatp-propositional-logic}\textbf{a}). As an illustrative example, consider a knowledge base with four propositional variables $A$, $B$, $C$, and $D$. The clause $A \vee \neg C$ is represented by the quantum state $\left| 1 \right\rangle_A \left| 0 \right\rangle_B \left| 2 \right\rangle_C \left| 0 \right\rangle_D$. In this encoding, $\left| 1 \right\rangle_A$ denotes the presence of the positive literal $A$; $\left| 0 \right\rangle_B$ and $\left| 0 \right\rangle_D$ represent the absence of $B$ and $D$ in the clause; and $\left| 2 \right\rangle_C$ corresponds to the negated literal $\neg C$. The remaining basis state, $\left| 3 \right\rangle$, is reserved to indicate the resolved status arising from the successful elimination of a pair of complementary literals during the resolution process.

We further define a quantum circuit within this framework to implement the resolution process in our QATP approach. 
The quantum resolution circuit takes three input ququarts. Two ququarts, prepared in the states $|c_1\rangle$ and $|c_2\rangle$, represent the premises of the resolution. These states encode the status (positive, negative, or absent) of a common propositional variable in their respective clauses. The third ququart is initialized in the state $|0\rangle$ and stores the resolution result produced by the circuit, as described by (Fig.~\ref{fig-qatp-propositional-logic}\textbf{c})
	\begin{align}\label{eq-qresolution}
		{U_\text{R}}\left| {{c_1}} \right\rangle \left| {{c_2}} \right\rangle \left| 0 \right\rangle  = \left| {{c_1}} \right\rangle \left| {{c_2}} \right\rangle \left| \mathcal{R}({c_1}, {c_2}) \right\rangle,
	\end{align}
where $\mathcal{R}({c_1}, {c_2})$ denotes the conclusion from resolving $c_1$ and $c_2$. The circuit begins by identifying complementary literals in the two input clauses. When such a pair is detected, the circuit marks their successful resolution by setting the corresponding ququart to the state $|3\rangle$. This state encodes the fact that the literals have been resolved, while all remaining ququarts preserve their previous values, reflecting the unaltered components of the clauses. Since this procedure can be executed in parallel across all propositional variables, the overall quantum circuit operates with a constant depth. By maintaining clause polarity and faithfully resolving complementary literals, the circuit accurately performs the quantum resolution operation, ensuring that the quantum-encoded logical structure undergoes the correct transformation. The detailed implementation of the quantum resolution unitary $U_\text{R}$ is provided in the Supplementary Information Fig.~1.

In the classical ATP scenario, for a knowledge base containing $M$ clauses, the resolution inference requires evaluating every possible pairwise combination of clauses. This results in a query complexity of $O(M^2)$ with respect to the knowledge base. In contrast, within our QATP framework, the unitary operator $U_\text{KB}$ (Fig.~\ref{fig-qatp-propositional-logic}\textbf{b}) is used to load the knowledge base into a quantum memory, after which the amplitude amplification subroutine (Fig.~\ref{fig-qatp-propositional-logic}\textbf{d}) increases the probability of observing valid resolution outcomes. Suppose that there are $s$ valid resolution results, where $s$ is an unknown but typically small integer. By employing the fixed-point quantum search algorithm~\cite{Yoder201411Fixed}, the QATP process requires only $O(M/\sqrt{s} )$ queries to $U_\text{KB}$. Considering that $O(s\log s)$  samples are needed to recover all the valid resulting clauses, the total query complexity becomes $O(M\sqrt{s} \log s )$, yielding a quadratic speedup over its classical counterpart. We note that such a quadratic speedup has the same origin as Grover's search algorithm \cite{Grover1996fast}. In addition, we mention that propositional entailment is co-NP complete \cite{Russell2020Artificial, Arora2009Computational},  hence every known inference algorithm for propositional logic has a worst-case complexity that is exponential in the size of the input and quantum algorithms cannot solve all cases efficiently.

The QATP framework can be extended to first-order logic in a conceptually straightforward manner. This extension proceeds by first transforming first-order sentences into clause form via Skolemization and conversion into CNF. A central part of this process is the elimination of existential quantifiers through Skolemization \cite{Robinson196501Machine}, which replaces each existential variable with a Skolem function or constant depending on the universally quantified variables in scope, while preserving satisfiability. The remaining universal quantifiers can then be removed, since variables in clause form are implicitly universally quantified. The resulting quantifier-free first-order clauses provide the basis for subsequent propositionalization.

After this transformation, first-order reasoning can in principle be carried out using resolution. However, resolution in first-order logic typically requires an additional unification step, which is computationally expensive on classical computers \cite{Russell2020Artificial} and poses significant challenges for direct quantum implementation. We circumvent this difficulty by invoking Herbrand’s theorem \cite{Russell2020Artificial}, which states that a set $W$ of clauses is unsatisfiable if and only if some finite subset of its Herbrand base $H_W(W)$ is unsatisfiable. Here, $H_W$ denotes the Herbrand universe of $W$, consisting of all ground terms generated from the function and constant symbols appearing in $W$. By working with ground instances drawn from the Herbrand universe, first-order reasoning is reduced to propositional reasoning, enabling the application of the QATP framework without requiring quantum unification.

    \begin{figure}[htb]
		\includegraphics[width=1\linewidth]{\figpath/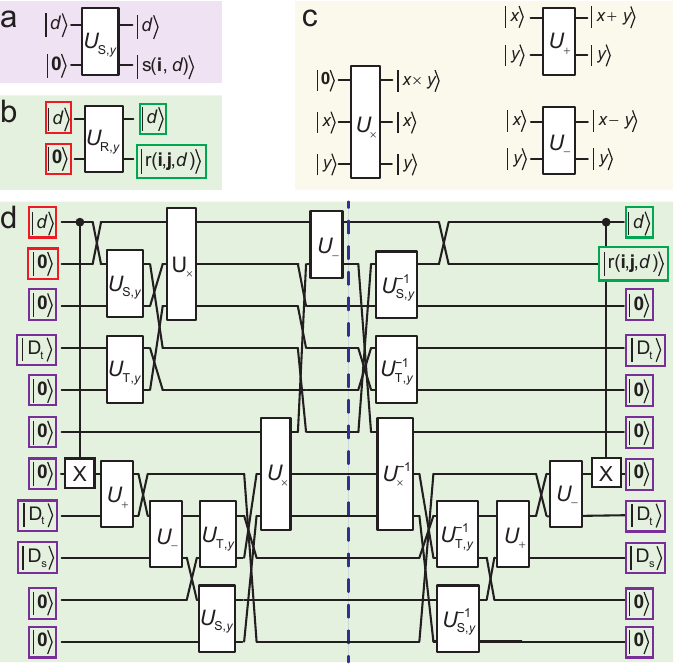}
		\caption{\label{fig-qp-unitary}\textbf{The quantum pseudo-division algorithm.} We consider a general scenario of dividing a polynomial $\text{S}$ by another polynomial $\text{T}$ with respect to the variable $y$. \textbf{a}, A sketch of ${U_{\text{S},y}}$, which represents the coefficients of $y$ with varying degrees. The qubits that encode the variables $\bf{x}$ are omitted, and the state $\left| {\sum\limits_{\bf{i}}^{\bf{I}} {\text{s}({\bf{i}},d)\prod\limits_k^\text{K} {x_k^{{i_k}}} } } \right\rangle $ is abbreviated as $\left| {\text{s}({\bf{i}},d)} \right\rangle $. \textbf{b}, A sketch of ${U_{\text{R},y}}$ that represents the remainder polynomial for $\text{S}$ divided by $\text{T}$. Here, for simplicity, we write $\left| {\text{r}({\bf{i}},{\bf{j}},d)} \right\rangle $ as the abbreviation of $\left| {\sum\limits_{{\bf{i}}, \bf{j}}^{\bf{I}, \bf{J}} {\text{r}({\bf{i}},{\bf{j}},d)\prod\limits_k^\text{K} {x_k^{{i_k} + {j_k}}} } } \right\rangle $. The number of qubits represented by $\left| \bf{0} \right\rangle$ in the red box is generally greater than the input state $\left| \bf{0} \right\rangle$ for ${U_{\text{R},y}}$ or ${U_{\text{T},y}}$, because these qubits for ${U_{\text{R},y}}$ represent the product of the polynomials for the coefficients of $\text{S}$ and $\text{T}$. The resulting polynomials usually have a higher degree. \textbf{c}, Basic arithmetic circuits used in the construction of ${U_{\text{R},y}}$. The circuits $U_{+}$, $U_{-}$, and $U_{\times}$ are designed to implement addition, subtraction, and multiplication of the input states, respectively.
        \textbf{d}, The quantum circuit for ${U_{\text{R},y}}$ that represents the remainder polynomial. The states in the red boxes are the input states and those in the green boxes are the output states of ${U_{\text{R},y}}$. The key output $\left| {{\rm{r}}({\bf{i}},{\bf{j}},d)} \right\rangle  = \left| {{\rm{s}}({\bf{i}},d){\rm{t}}({\bf{j}},{{\rm{D}}_{\rm{t}}}) - {\rm{t}}({\bf{j}},d + {{\rm{D}}_{\rm{t}}} - {{\rm{D}}_{\rm{s}}}){\rm{s}}({\bf{i}},{{\rm{D}}_{\rm{s}}})} \right\rangle $ is obtained after the portion of the circuit preceding the blue dashed line is executed. The remaining qubits in the purple boxes serve as auxiliary qubits that can be prepared to the desired states (purple boxes) with one layer of single-qubit gates. These auxiliary qubits are reset after the state $\left| {\text{r}({\bf{i}},{\bf{j}},d)} \right\rangle $ is produced. }
    \end{figure}

\vspace{.5cm}
\noindent\textbf{\large{}QATP for geometric theorems}{\large\par} 

\noindent We now discuss QATP for geometric theorems, with a focus on Wu's method, which is a powerful and seminal algebraic approach to automated geometric theorem proving~\cite{Chou198809introduction}. Instead of reasoning on geometric figures directly, Wu’s method encodes geometric conditions—such as incidences, parallels, or perpendicularities—into systems of polynomial equations defined over coordinate variables. The central task then becomes determining whether a set of hypothesis polynomials implies a conclusion polynomial. The core mechanism of Wu’s method involves polynomial pseudo-division and successive algebraic elimination, which systematically transform and reduce the polynomial system to a canonical form. 
If the final remainder polynomial vanishes identically after a sequence of reductions, the geometric theorem is thereby proved. Wu’s method has exerted a lasting influence on computer algebra and symbolic reasoning, forming the foundation of several modern systems for automated geometric theorem proving and computer-assisted proof verification. \cite{Chou1994MachineProofs}. However, its computational complexity—dominated by high-degree polynomial manipulations—poses significant scalability challenges, motivating the exploration of potential quantum-enhanced implementations.

   \begin{figure*}[htbp]
        \centering
	\includegraphics[width=1\linewidth]{\figpath/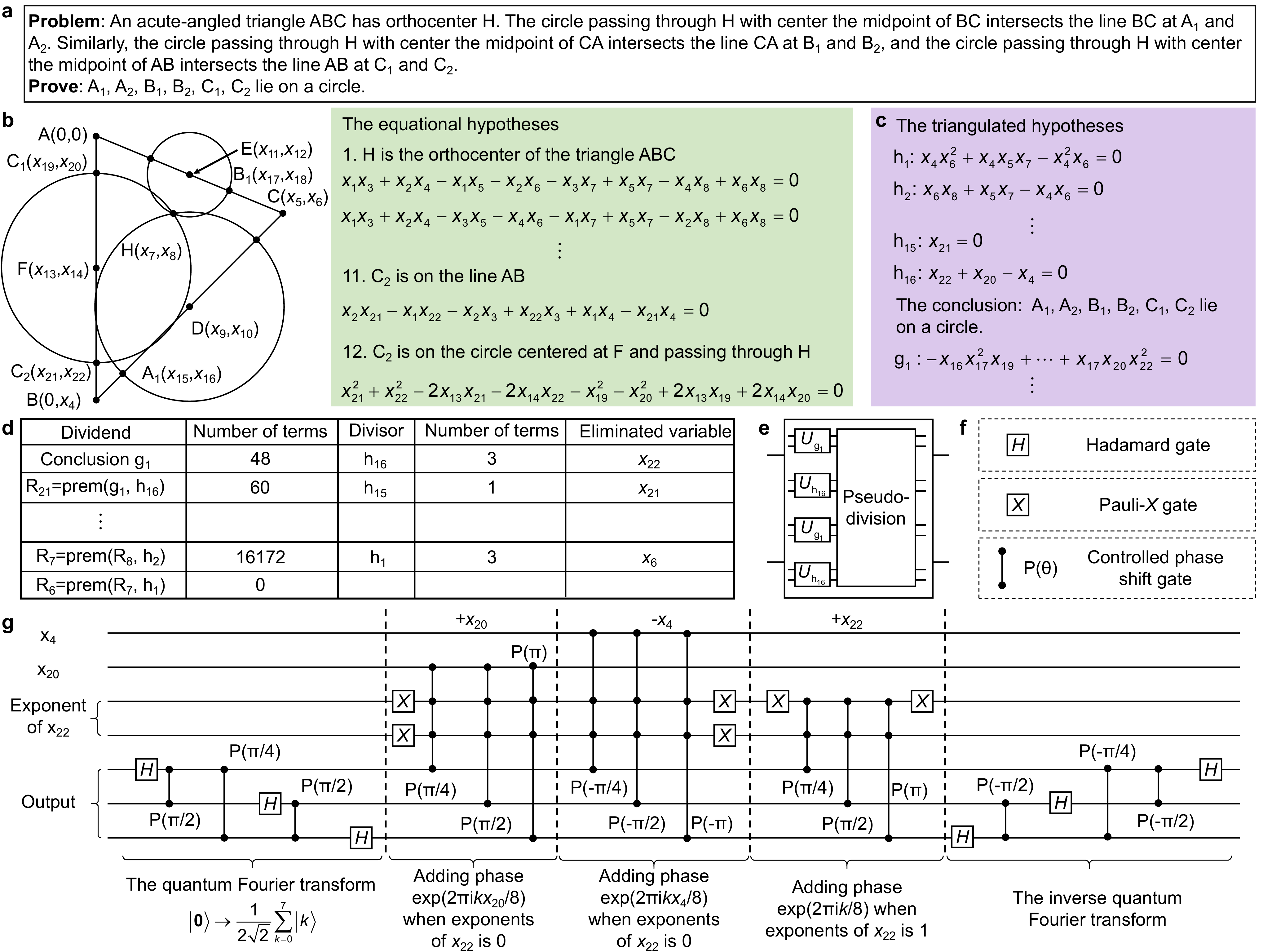}
	\caption{\label{fig-qpp-2008p1b} \textbf{An illustration of QATP for an IMO geometry problem.}
    \textbf{a}, The IMO 2008 Geometry Problem 1 in natural language. \textbf{b}, Coordinate assignment and hypothesis algebraic equations for this geometry problem. Without loss of generality, we assign the coordinates of the first point A as (0, 0), and those of the second point B as (0, $x_4$). This choice corresponds to setting $x_1 = x_2 = x_3 = 0$. For simplicity, here we only explicitly show a small portion of the hypothesis algebraic equations.
    \textbf{c}, The triangular form of the hypothesis equations and one of the conclusion equation $g_1$.
    \textbf{d}, Successive pseudo-division of the conclusion by the triangulated hypotheses. The variable with the largest subscript in the conclusion is $x_{22}$, which is removed in the first pseudo-division using $h_{16}$. The last pseudo-remainder $R_6$ is zero, proving the validity of the conclusion $g_1$. The number of monomials in the pseudo-remainders determines the computational time and space required for a given proof, potentially being very large for certain  cases \cite{Wu1978Decision,Chou198809introduction}. For this particular problem, the largest number is 16172.
    \textbf{e}, A sketch of the quantum circuit for pseudo-division of the conclusion polynomial $g_1$ by $h_{16}$. This circuit involves  two copies of the circuit representing $g_1$ and two copies of the circuit representing $h_{16}$, together with an arithmetic circuit required to perform pseudo-division.
    \textbf{f}, Quantum gates used in representing polynomials with quantum circuits.
    \textbf{g}, The quantum circuit for representing $h_{16}$ with seven qubits. This circuit calculates the coefficients of $h_{16}$ in variable $x_{22}$. We first perform quantum Fourier transform on the output qubits, and then use controlled phase gates to add phases determined by the value of the input variables and the degree of the variable being eliminated. The last step is to revert the phase of the quantum state back to the binary representation on the qubits. We use the circuit for $h_{16}$ on seven qubits in this figure as a simple demonstration. The complete circuits of $h_{16}$ employed in the first step of quantum pseudo-division in fact requires more qubits to represent $x_4$, $x_{20}$, the exponent of $x_{22}$, and the output. It involves 28 qubits and has a depth of 89, which is too complicated for clear visualization.}
\end{figure*}

To extend Wu's method into the quantum domain, a crucial step is to implement polynomial pseudo-division on a quantum computer. To achieve this, we first need to find an appropriate representation for encoding polynomials in quantum form. A straightforward approach exploits quantum state encoding of polynomial coefficients. The polynomial is represented as an entangled quantum state, with each component corresponding to a single coefficient (see Supplementary  Information Section III.B for details).
However, this approach faces a number of challenges. First, polynomial multiplication and subtraction required in pseudo-division involve arithmetic operations on coefficients. Although basic integer arithmetic can be efficiently performed using quantum Fourier transform–based algorithms~\cite{RuizPerez201704Quantum}, the task of combining like terms in polynomial multiplication remains cumbersome. This difficulty arises because, under the quantum superposition principle, the coefficients of a polynomial are encoded across multiple superposed states on the same qubits. Comparing degrees and merging coefficients of equal degree fall beyond the scope of standard quantum arithmetic algorithms~\cite{Takahashi2009Quantum}. Second, combining like terms is inherently irreversible, further complicating its quantum implementation. Third, the quantum no-cloning theorem~\cite{Wootters1982Single} forbids the duplication of arbitrary unknown quantum states, preventing the direct reuse of the resulting quantum state from one pseudo-division step as input to the next. Finally, after each pseudo-division step, the resulting state becomes highly entangled with the input polynomial states, complicating subsequent operations and rendering the overall process highly intricate and resource-intensive.

We overcome these difficulties by introducing a quantum pseudo-division algorithm based on circuit encoding of the point-value representation of polynomials (Fig.~\ref{fig-qp-unitary}). In such a representation, a polynomial $\text{F}(\mathbf{x})$ is represented by $M$ point-value pairs $\{\mathbf{x},\text{F}(\mathbf{x})\}$, where $\mathbf{x}=(x_1,x_2,\cdots, x_\text{K})$ denotes the variables. With this representation, algebraic operations of polynomials reduce to operations on corresponding point-value pairs, thus circumventing the first and second challenges faced by the quantum state encoding approach discussed above. In the quantum scenario, these $M$ point-value pairs are further encoded into a quantum state $|\psi {\rangle _{\rm{F}}} = \sum\nolimits_m^M {\left| {{{\bf{x}}_m}} \right\rangle \left| {{\rm{F}}({{\bf{x}}_m})} \right\rangle } $, which in turn can be prepared by a unitary $U_\text{F}$ specified by a quantum circuit with a depth scaling linearly with $M$:
\begin{eqnarray} 
		{U_\text{F}}\left| {\bf{x}} \right\rangle {\left|\bf{0}\right\rangle} = \left| {\bf{x}} \right\rangle \left| {\text{F}({\bf{x}})} \right\rangle.
\end{eqnarray}
In this way, each polynomial is represented by a unitary implemented by a quantum circuit. This representation allows cloning of the polynomial, and reduces pseudo-division to the construction of the quantum circuit for the remainder polynomial (Fig.~\ref{fig-qp-unitary}\textbf{b,d}). More concretely, we consider dividing polynomial ${\rm{S}} = \sum\limits_{{\bf{i}},{d_1}}^{{\bf{I}},{{\rm{D}}_{\rm{s}}}} {{\rm{s}}({\bf{i}},{d_1})\prod\limits_k^{\rm{K}} {x_k^{{i_k}}} {y^{{d_1}}}} $ by ${\rm{T}} = \sum\limits_{{\bf{j}},{d_2}}^{{\bf{J}},{{\rm{D}}_{\rm{t}}}} {{\rm{t}}\left( {{\bf{j}},{d_2}} \right)\prod\limits_k^{\rm{K}} {x_k^{{j_k}}} {y^{{d_2}}}} $ with respect to the variable $y$. Here, $\text{s}({\bf{i}},{d_1})$ and $\text{t}({{\bf{j}},{d_2}})$ represent the coefficients with $\mathbf{i}\equiv (i_1,i_2,\cdots,i_\text{K})$ and $\mathbf{j}\equiv (j_1,j_2,\cdots,j_\text{K})$ denoting collectively the exponents of corresponding variables;  the maximum degrees of these two polynomials in the variables $x_k$ and $y$ are denoted as $\text{I}_k$, $\text{J}_k$, $\text{D}_\text{s}$, and $\text{D}_\text{t}$, respectively. Without loss of generality, we suppose ${{\rm{D}}_{\rm{s}}} > {{\rm{D}}_{\rm{t}}}$. In order to obtain the remainder polynomial for $\text{S}$ divided by $\text{T}$ with respect to $y$, a crucial subroutine involves computing the coefficient of variable $y$ with varying degrees. Those coefficients are themselves polynomials in variables $\bf{x}$, which in turn can be represented by a unitary $U_{\text{S},y}$ (Fig.~\ref{fig-qp-unitary}\textbf{a}):    
    \begin{eqnarray}\label{eq-usy}
        {U_{\text{S},y}}\left| d \right\rangle \left| {\bf{x}} \right\rangle \left| {\bf{0}} \right\rangle  = \left| d \right\rangle \left| {\bf{x}} \right\rangle \left| {\sum\limits_{\bf{i}}^{\bf{I}} {\text{s}({\bf{i}},d)\prod\limits_k^\text{K} {x_k^{{i_k}}} } } \right\rangle.
    \end{eqnarray}
Here, $\sum\limits_{\bf{i}}^{\bf{I}} {\text{s}({\bf{i}},{d})\prod\limits_k^\text{K} {x_k^{{i_k}}} } $  denotes the coefficient of $y^d$ in $\text{S}$. An illustration of our quantum pseudo-division algorithm is shown in Fig.~\ref{fig-qp-unitary}. With this quantum pseudo-division algorithm, the general procedure for proving geometric theorems with a quantum computer is now straightforward and consists of the following major steps: Step $1$, for a given geometric theorem proving problem,  we translate it into a series  of algebraic equations, yielding a set of hypothesis equations and another set of conclusion equations; Step $2$, we transform the set of hypotheses equations into the triangular form by using the quantum pseudo-division algorithm repeatedly; Step $3$, for each conclusion equation, we perform successive pseudo-division on the transformed hypotheses in triangular form to obtain the final remainder; Step $4$, we measure the final state to check whether the final remainder is zero. If it is, the corresponding conclusion is then proved. We mention that step $2$ and $3$ can be implemented with a unified quantum circuit and one only needs to measure at the step $4$ (Supplementary Information Section III.C).

As an illustrative example, we now apply our general protocol for proving geometric theorems to the IMO 2008 Geometry Problem 1 (Fig.~\ref{fig-qpp-2008p1b}\textbf{a}) \cite{IMOOfficialWebsite,Trinh202401Solving, Sinha2024Wus}.
In Fig.~\ref{fig-qpp-2008p1b}\textbf{b}, we present a diagram for the IMO problem in the coordinate plane and translate the hypotheses of the problem into algebraic equations. We then triangulate these hypotheses through pseudo-divisions so that each subsequent hypothesis equation introduces only one new variable (Fig.~\ref{fig-qpp-2008p1b}\textbf{c}). We prove the conclusion by successively eliminating variables through pseudo-division, starting with the variable with the largest subscript and the last hypothesis polynomial. More concretely, we start by dividing the conclusion polynomial $g_1$ by $h_{16}$ with respect to $x_{22}$, leading to a pseudo-remainder which is further divided by $h_{15}$ to remove $x_{21}$. We repeat these steps and find that the final pseudo-remainder $R_6$ for dividing $R_7$ by $h_1$ with respect to $x_6$ is zero (Fig.~\ref{fig-qpp-2008p1b}\textbf{d}), thus proving the conclusion $g_1$.

The key subroutine for executing quantum pseudo-division is implemented by constructing the pseudo-remainder circuit from circuits representing the dividend and divisor, as illustrated in Fig.~\ref{fig-qpp-2008p1b}\textbf{e}. The resulting polynomial is represented by the circuit $U_\text{R}$, which incorporates two copies of the circuit corresponding to the conclusion $g_1$, two copies of the circuit for the divisor $h_{16}$, and additional arithmetic subcircuits that realize the pseudo-division operations as shown in Fig. \ref{fig-qp-unitary}\textbf{d}. For the demonstration of the quantum implementation of pseudo-division, we present a scheme for constructing a circuit that represents a multivariate polynomial as a univariate polynomial with respect to a chosen variable ($x_{22}$ in this case), where the circuit depth scales linearly with the number of nonzero terms. A general framework for circuit construction is provided later in this section.
As shown in Fig.~\ref{fig-qpp-2008p1b}\textbf{g}, we illustrate the construction for the polynomial $h_{16}$ implemented on seven qubits. Since $h_{16}$ serves as the divisor for eliminating the variable $x_{22}$, it can be expressed as
$h_{16} = 1 \cdot x_{22}^1 + (x_{20} - x_4) \cdot x_{22}^0$. In this circuit, the first two qubits encode the respective values of $x_4$ and $x_{20}$, while the next two qubits represent the exponent of $x_{22}$. Although the exponent of $x_{22}$ in $h_{16}$ is either 0 or 1, two qubits are required to accommodate the binary representation of ``2'' (i.e., ``10''), as $h_{16}$ is multiplied by $x_{22}$ during the division of the conclusion $g_1$. The remaining three qubits encode the coefficient values of $h_{16}$ in $x_{22}$ for different exponents, together with the input values of $x_4$ and $x_{20}$.

We apply a quantum Fourier transform to the three output qubits initially in the state $\left| \mathbf{0} \right\rangle$, producing the uniform superposition $\frac{1}{\sqrt{8}} \sum_{k=0}^{7} \left| k \right\rangle$. To compute all monomials in $h_{16}$, controlled phase-shift gates are applied to the output qubits, conditioned on the qubits encoding $x_{20}$, $x_4$, and the exponent of $x_{22}$. After executing all controlled operations, the output qubits evolve to the state $\sum_{k} \frac{e^{\pi i k (x_{20} - x_4)/4}}{2\sqrt{2}} \left| k \right\rangle$ if the exponent of $x_{22}$ is 0, or to $\sum_{k} \frac{e^{\pi i k /4}}{2\sqrt{2}} \left| k \right\rangle$
if the exponent is 1. Subsequently, an inverse quantum Fourier transform is applied to recover the binary-encoded coefficients of $h_{16}$ as functions of $x_{22}$. The complete circuits used for the first step of quantum pseudo-division employ 28 qubits, with circuit depths of 4068 for the conclusion $g_1$ and 89 for $h_{16}$. Notably, in each quantum pseudo-division step, the resulting circuit is constructed directly from those of the dividend and divisor (as depicted in Fig.~\ref{fig-qpp-2008p1b}\textbf{e}), implying that its depth is independent of the number of monomials. This contrasts with the classical implementation of Wu’s method that operates on  coefficient representations of polynomials \cite{Ye2008introduction}, where both space and time complexity increase sharply as the number of monomials grows across successive pseudo-remainders during the proof process.

We now analyze the complexity of implementing Wu’s method on a quantum computer. We focus on two aspects: (i) the gate complexity required to encode multivariate polynomials and perform arithmetic operations on them, and (ii) the query complexity associated with polynomial identity testing (PIT) \cite{Shpilka201012Arithmetic, Saxena2009Progress, Saxena2014Progress}, namely determining whether the final pseudo-remainder is identically zero or not. As discussed above, within the framework for quantum implementation of geometric theorem proving, the overall workflow decomposes into three fundamental subroutines, whose complexities are analyzed one-by-one in the following.

The first subroutine concerns the construction of quantum circuits that encode the hypothesis polynomials. We outline two strategies for this task. One approach is to classically evaluate a polynomial at $M$ input points, thereby generating $M$ point–value pairs, which are subsequently loaded into a quantum circuit. This yields a circuit depth of $O(M\log M)$, analogous to the knowledge-base encoding unitary $U_\text{KB}$ shown in Fig.~\ref{fig-qatp-propositional-logic}\textbf{b}. Here, $M$ is determined by the number of evaluation points required in the final PIT stage. An alternative strategy is to construct the polynomial circuit directly from its algebraic structure, explicitly encoding each monomial and summing them coherently. As illustrated in Fig.~\ref{fig-qpp-2008p1b}\textbf{g}, a polynomial with $M$ nonzero monomials gives rise to a circuit whose depth scales linearly with $M$. The gate complexity associated with generating a single monomial is analyzed in the Supplementary Information Section III.C. This direct construction allows full freedom in selecting the number of point–value evaluations used later in PIT. Importantly, the overhead of building the hypothesis circuits is modest, as this step is performed only once at the outset of the proof protocol. Moreover, in typical geometric reasoning tasks, the hypothesis polynomials are substantially simpler than the intermediate polynomials generated during subsequent proof derivations.

The second subroutine constructs a circuit that represents a polynomial with respect to a chosen division variable. This task can be formulated as an interpolation problem: given a circuit $U_\text{S}$ that evaluates a multivariate polynomial $\text{S}$, the objective is to construct a circuit 
$U_{\text{S},y}$ that outputs the coefficients of $\text{S}$ corresponding to different powers of the variable $y$. Classical algorithms typically implement this step using the fast Fourier transform (FFT)~\cite{Cooley1965algorithm, Schoenhage197109Schnelle, Borodin197406Fast}. While such reversible classical procedures can, in principle, be translated into quantum circuits~\cite{Nielsen201206Quantum}, we instead adopt Kravchuk polynomials~\cite{Szegoe1939Orthogonal} as the interpolation basis.
This choice avoids the substantial overhead associated with complex-valued floating-point arithmetic in FFT-based methods. The resulting construction requires ${{\rm{D}}_{\rm{s}}}$ queries to $U_\text{S}$, matching the query complexity of the classical FFT approach in polynomial interpolation, while the depth of the remaining operations scales as $O({{\rm{D}}_{\rm{s}}}\log {{\rm{D}}_{\rm{s}}})$ when suitable ancillary qubits are available. Additional details are provided in the Supplementary Information Section III.C.
The third subroutine comprises the quantum arithmetic circuits required to implement pseudo-division. This component consists of integer addition, subtraction, and multiplication modules, as illustrated in Fig.~\ref{fig-qp-unitary}\textbf{d}. The circuit depth of these arithmetic operations scales polynomially with the number of qubits involved~\cite{RuizPerez201704Quantum}. In practice, this cost is insignificant compared with the depth of the circuits encoding the polynomials themselves. In particular, encoding $M$ evaluation points requires only $\log_2 M$ qubits to represent the evaluation domain.

After constructing the circuit that computes the final pseudo-remainder, the remaining task is to determine whether this circuit represents the zero polynomial. This is achieved by evaluating the polynomial on a finite set $\text{G}$ of points and checking whether it vanishes on all elements of $\text{G}$. A fixed-point quantum search algorithm~\cite{Yoder201411Fixed} can be employed to search for points at which the polynomial evaluates to a nonzero value. Specifically, if there are $h$ points in $\text{G}$ at which the polynomial is nonzero, then $O(\log (2/\delta )\sqrt {|\text{G}|/h} )$ queries to the quantum evaluation circuit suffice to achieve a success probability of
$1-\delta^2$. By contrast, a classical randomized strategy based on independent sampling with replacement requires $O(\log (1/\delta)|\text{G}|/h)$ queries to reach the same confidence level in the regime where $h$ is small. This yields a quadratic speedup with respect to the size of the evaluation set $|\text{G}|$.

\vspace{.5cm}
\noindent\textbf{\large{}Discussion and outlook}{\large\par}    

\noindent In summary, we have introduced a general QATP framework that applies quantum computation to accelerate core inference procedures in classical automated reasoning. By developing quantum implementations of the resolution method for propositional and first-order logic, as well as a quantum formulation of Wu’s method for geometric theorem proving, we have shown that established symbolic reasoning paradigms can be embedded into quantum computational paradigms without modifying their logical foundations.

For propositional and first-order logic, our quantum resolution procedure achieves a quadratic speedup in query complexity for identifying valid resolvents by replacing classical exhaustive search with quantum search. This speedup is optimal for unstructured search problems and aligns with known lower bounds in the quantum query model~\cite{Beals200107Quantum, Aaronson2014Need}. For geometric theorem proving, we reformulated Wu’s method as an quantum algebraic reasoning process based on polynomial pseudo-division, reducing proof verification to polynomial identity testing. Quantum circuits enable the coherent evaluation of polynomials at multiple points via superposition, and quantum search yields an overall quadratic advantage in detecting nonzero evaluations. A circuit realization for the 2008 IMO Geometry Problem 1 demonstrates the feasibility of end-to-end geometric proofs within this framework.

More broadly, our work represents an primary step toward the systematic adaptation of automated theorem proving to the quantum domain. The quantum reasoning procedures introduced here evaluate the validity of a conclusion given access to a knowledge base or a set of geometric relations and can be rigorously analyzed within the quantum decision-tree (query) model. In this setting, QATP naturally appears as a search problem—for example, searching for a contradiction in resolution or for nonzero evaluations in PIT—while also fitting into the broader framework of total Boolean function evaluation~\cite{Buhrman200210Complexity, Aaronson2014Need}. For such functions, the bounded-error quantum query complexity is lower bounded by the sixth root of the classical deterministic query complexity~\cite{Beals200107Quantum, Aaronson2014Need}. Within this theoretical landscape, our approach achieves a quadratic separation, matching the largest separation currently known for unstructured problems and enabled by quantum search techniques~\cite{Grover1996fast, Bennett199710Strengths}.

We also investigated quantum circuit constructions for representing and manipulating multivariate polynomials, which are central to algebraic methods for geometric reasoning. Although the quantum circuit model is universal~\cite{Deutsch1985Quantum, Bernstein1993Quantum, ChiChihYao1993Quantum}, implementing an arbitrary $n$-bit reversible function can still require $\Omega(n 2^n / \log n)$ elementary gates in the worst case~\cite{Shende200306Synthesis, Wu202412Asymptotically}. Our focus therefore lies on the structured algebraic operations arising in Wu’s method. In particular, we developed quantum circuit constructions for polynomial pseudo-division and employed Kravchuk polynomials as an interpolation basis over integer-valued domains, thereby avoiding costly complex-valued floating-point arithmetic and enabling efficient circuit realizations. These techniques allow multivariate polynomials to be manipulated coherently and integrated into end-to-end quantum geometric proofs.

While evaluating polynomial values may appear straightforward, the algebraic transformations required in geometric theorem proving—such as pseudo-division, interpolation, and successive elimination—introduce substantial structural complexity. Understanding how these structured operations interact with quantum resources remains an open problem. An important direction for future work is to identify restricted classes of geometric constraints whose algebraic structure may admit stronger quantum advantages, potentially exceeding quadratic speedups.
Looking forward, it would be interesting and important to carry out experimental demonstrations of quantum automated reasoning with current noisy or early fault-tolerant quantum devices. This would be a crucial step toward practical applications of quantum technologies in symbolic artificial intelligence.

	\vspace{.5cm}
	\noindent\textbf{Acknowledgements} We thank X. Gao, A. Anshu, P. H. Yao, X. Sun, W. J. Jiang, W. K. Li, X.-S. Gao, M. Ying, and L.-M. Duan for enlightening discussions. This work is supported by the National Natural Science Foundation of China (Grant No. T2225008 and T24B2002), the Innovation Program for Quantum Science and Technology (Grant No. 2021ZD0302203), the Tsinghua University Dushi Program, and the Shanghai Qi Zhi Institute Innovation Program SQZ202318. Z.-Z. S. is additionally funded by China Postdoctoral Science Foundation (Certificated Number: 2025T180926).
    
\let\oldaddcontentsline\addcontentsline
\renewcommand{\addcontentsline}[3]{}

\bibliography{ref.bib}

@Article{RuizPerez201704Quantum,
  author    = {Lidia Ruiz-Perez and Juan Carlos Garcia-Escartin},
  journal   = {Quantum Inf. Process.},
  title     = {Quantum arithmetic with the quantum Fourier transform},
  year      = {2017},
  month     = {apr},
  number    = {6},
  volume    = {16},
pages={152},
  doi       = {10.1007/s11128-017-1603-1},
  publisher = {Springer Science and Business Media {LLC}},
}

@Article{Chou198809introduction,
  author    = {Shang-Ching Chou},
  journal   = {J. Autom. Reasoning},
  title     = {An introduction to {W}u{\textquotesingle}s method for mechanical theorem proving in geometry},
  year      = {1988},
  month     = {sep},
  number    = {3},
  volume    = {4},
  pages = {237-267},
  doi       = {10.1007/bf00244942},
  publisher = {Springer Science and Business Media {LLC}},
}

@article{Gowers2025Conjecture,
  title={On a conjecture of Marton},
  author={Gowers, William Timothy and Green, Ben and Manners, Freddie and Tao, Terence},
  journal={Ann. of Math.},
  volume={201},
  number={2},
  pages={515--549},
  year={2025},
  publisher={Department of Mathematics, Princeton University Princeton, New Jersey, USA},
url={https://doi.org/10.4007/annals.2025.201.2.5},
doi={10.4007/annals.2025.201.2.5}
}

@Article{Trinh202401Solving,
  author    = {Trinh, Trieu H. and Wu, Yuhuai and Le, Quoc V. and He, He and Luong, Thang},
  journal   = {Nature},
  title     = {Solving olympiad geometry without human demonstrations},
  year      = {2024},
  issn      = {1476-4687},
  month     = jan,
  number    = {7995},
  pages     = {476--482},
  volume    = {625},
  doi       = {10.1038/s41586-023-06747-5},
  publisher = {Springer Science and Business Media LLC},
}

@Article{Arute201910Quantum,
  author    = {Arute, Frank and Arya, Kunal and Babbush, Ryan and Bacon, Dave and Bardin, Joseph C. and Barends, Rami and Biswas, Rupak and Boixo, Sergio and Brandao, Fernando G. S. L. and Buell, David A. and others},
  journal   = {Nature},
  title     = {Quantum supremacy using a programmable superconducting processor},
  year      = {2019},
  issn      = {1476-4687},
  month     = oct,
  number    = {7779},
  pages     = {505--510},
  volume    = {574},
  doi       = {10.1038/s41586-019-1666-5},
  publisher = {Springer Science and Business Media LLC},
}

@Article{Wu202110Strong,
  author    = {Wu, Yulin and Bao, Wan-Su and Cao, Sirui and Chen, Fusheng and Chen, Ming-Cheng and Chen, Xiawei and Chung, Tung-Hsun and Deng, Hui and Du, Yajie and Fan, Daojin and others},
  title     = {Strong Quantum Computational Advantage Using a Superconducting Quantum Processor},
  year      = {2021},
  issn      = {1079-7114},
  month     = oct,
  number    = {18},
  pages     = {180501},
  volume    = {127},
  doi       = {10.1103/PhysRevLett.127.180501},
  publisher = {American Physical Society (APS)},
  journal = prl,
}

@Book{Nielsen201206Quantum,
  author    = {Nielsen, Michael A. and Chuang, Isaac L.},
  title     = {Quantum Computation and Quantum Information: 10th Anniversary Edition},
  year      = {2012},
  month     = jun,
  isbn      = {9780511976667},
  publisher = {Cambridge University Press},
}

@Article{Zhong202012Quantum,
  author    = {Zhong, Han-Sen and Wang, Hui and Deng, Yu-Hao and Chen, Ming-Cheng and Peng, Li-Chao and Luo, Yi-Han and Qin, Jian and Wu, Dian and Ding, Xing and Hu, Yi and others},
  journal   = {Science},
  title     = {Quantum computational advantage using photons},
  year      = {2020},
  issn      = {1095-9203},
  month     = dec,
  number    = {6523},
  pages     = {1460--1463},
  volume    = {370},
  doi       = {10.1126/science.abe8770},
  publisher = {American Association for the Advancement of Science (AAAS)},
}

@Article{Krinner202205Realizing,
  author    = {Krinner, Sebastian and Lacroix, Nathan and Remm, Ants and Di Paolo, Agustin and Genois, Elie and Leroux, Catherine and Hellings, Christoph and Lazar, Stefania and Swiadek, Francois and Herrmann, Johannes and others},
  journal   = {Nature},
  title     = {Realizing repeated quantum error correction in a distance-three surface code},
  year      = {2022},
  issn      = {1476-4687},
  month     = may,
  number    = {7911},
  pages     = {669--674},
  volume    = {605},
  doi       = {10.1038/s41586-022-04566-8},
  publisher = {Springer Science and Business Media LLC},
}

@Article{Kim202306Evidence,
  author    = {Kim, Youngseok and Eddins, Andrew and Anand, Sajant and Wei, Ken Xuan and van den Berg, Ewout and Rosenblatt, Sami and Nayfeh, Hasan and Wu, Yantao and Zaletel, Michael and Temme, Kristan and Kandala, Abhinav},
  journal   = {Nature},
  title     = {Evidence for the utility of quantum computing before fault tolerance},
  year      = {2023},
  issn      = {1476-4687},
  month     = jun,
  number    = {7965},
  pages     = {500--505},
  volume    = {618},
  doi       = {10.1038/s41586-023-06096-3},
  publisher = {Springer Science and Business Media LLC},
}

@Article{Ebadi202107Quantum,
  author    = {Ebadi, Sepehr and Wang, Tout T. and Levine, Harry and Keesling, Alexander and Semeghini, Giulia and Omran, Ahmed and Bluvstein, Dolev and Samajdar, Rhine and Pichler, Hannes and Ho, Wen Wei and others},
  journal   = {Nature},
  title     = {Quantum phases of matter on a 256-atom programmable quantum simulator},
  year      = {2021},
  issn      = {1476-4687},
  month     = jul,
  number    = {7866},
  pages     = {227--232},
  volume    = {595},
  doi       = {10.1038/s41586-021-03582-4},
  publisher = {Springer Science and Business Media LLC},
}

@Article{Madsen202206Quantum,
  author    = {Madsen, Lars S. and Laudenbach, Fabian and Askarani, Mohsen Falamarzi. and Rortais, Fabien and Vincent, Trevor and Bulmer, Jacob F. F. and Miatto, Filippo M. and Neuhaus, Leonhard and Helt, Lukas G. and Collins, Matthew J. and others},
  journal   = {Nature},
  title     = {Quantum computational advantage with a programmable photonic processor},
  year      = {2022},
  issn      = {1476-4687},
  month     = jun,
  number    = {7912},
  pages     = {75--81},
  volume    = {606},
  doi       = {10.1038/s41586-022-04725-x},
  publisher = {Springer Science and Business Media LLC},
}

@Article{Egan202110Fault,
  author    = {Egan, Laird and Debroy, Dripto M. and Noel, Crystal and Risinger, Andrew and Zhu, Daiwei and Biswas, Debopriyo and Newman, Michael and Li, Muyuan and Brown, Kenneth R. and Cetina, Marko and Monroe, Christopher},
  journal   = {Nature},
  title     = {Fault-tolerant control of an error-corrected qubit},
  year      = {2021},
  issn      = {1476-4687},
  month     = oct,
  number    = {7880},
  pages     = {281--286},
  volume    = {598},
  doi       = {10.1038/s41586-021-03928-y},
  publisher = {Springer Science and Business Media LLC},
}

@Article{Moses202312Race,
  title = {A Race-Track Trapped-Ion Quantum Processor},
  author = {Moses, S. A. and Baldwin, C. H. and Allman, M. S. and Ancona, R. and Ascarrunz, L. and Barnes, C. and Bartolotta, J. and Bjork, B. and Blanchard, P. and Bohn, M. and others},
  journal = {Phys. Rev. X},
  volume = {13},
  issue = {4},
  pages = {041052},
  numpages = {25},
  year = {2023},
  month = {Dec},
  publisher = {American Physical Society},
  doi = {10.1103/PhysRevX.13.041052},
  url = {https://link.aps.org/doi/10.1103/PhysRevX.13.041052}
}

@Article{Acharya202302Suppressing,
  author    = {Acharya, Rajeev and Aleiner, Igor and Allen, Richard and Andersen, Trond I. and Ansmann, Markus and Arute, Frank and Arya, Kunal and Asfaw, Abraham and Atalaya, Juan and Babbush, Ryan and others},
  journal   = {Nature},
  title     = {Suppressing quantum errors by scaling a surface code logical qubit},
  year      = {2023},
  issn      = {1476-4687},
  month     = feb,
  number    = {7949},
  pages     = {676--681},
  volume    = {614},
  doi       = {10.1038/s41586-022-05434-1},
  publisher = {Springer Science and Business Media LLC},
}

@Article{Ni202303Beating,
  author    = {Ni, Zhongchu and Li, Sai and Deng, Xiaowei and Cai, Yanyan and Zhang, Libo and Wang, Weiting and Yang, Zhen-Biao and Yu, Haifeng and Yan, Fei and Liu, Song and others},
  journal   = {Nature},
  title     = {Beating the break-even point with a discrete-variable-encoded logical qubit},
  year      = {2023},
  issn      = {1476-4687},
  month     = mar,
  number    = {7955},
  pages     = {56--60},
  volume    = {616},
  doi       = {10.1038/s41586-023-05784-4},
  publisher = {Springer Science and Business Media LLC},
}

@Article{Bluvstein2024Logical,
  author    = {Bluvstein, Dolev and Evered, Simon J. and Geim, Alexandra A. and Li, Sophie H. and Zhou, Hengyun and Manovitz, Tom and Ebadi, Sepehr and Cain, Madelyn and Kalinowski, Marcin and Hangleiter, Dominik and others},
  journal   = {Nature},
  title     = {Logical quantum processor based on reconfigurable atom arrays},
  year      = {2024},
  issn      = {1476-4687},
  number    = {7997},
  pages     = {58--65},
  volume    = {626},
  doi       = {10.1038/s41586-023-06927-3},
  publisher = {Springer Science and Business Media LLC},
}

@Article{Sivak202303Real,
  author    = {Sivak, V. V. and Eickbusch, A. and Royer, B. and Singh, S. and Tsioutsios, I. and Ganjam, S. and Miano, A. and Brock, B. L. and Ding, A. Z. and Frunzio, L. and others},
  journal   = {Nature},
  title     = {Real-time quantum error correction beyond break-even},
  year      = {2023},
  issn      = {1476-4687},
  month     = mar,
  number    = {7955},
  pages     = {50--55},
  volume    = {616},
  doi       = {10.1038/s41586-023-05782-6},
  publisher = {Springer Science and Business Media LLC},
}

@Article{Xu202407Non,
  author    = {Xu, Shibo and Sun, Zheng-Zhi and Wang, Ke and Li, Hekang and Zhu, Zitian and Dong, Hang and Deng, Jinfeng and Zhang, Xu and Chen, Jiachen and Wu, Yaozu and others},
  journal   = {Nat. Phys.},
  title     = {Non-Abelian braiding of Fibonacci anyons with a superconducting processor},
  year      = {2024},
  issn      = {1745-2481},
  month     = jul,
  number    = {9},
  pages     = {1469--1475},
  volume    = {20},
  doi       = {10.1038/s41567-024-02529-6},
}

@Book{Russell2020Artificial,
  author    = {Russell, Stuart J. and Norvig, Peter},
  publisher = {Pearson},
  title     = {Artificial Intelligence A Modern Approach},
  year      = {2020},
  isbn      = {9780134610993},
  subtitle  = {A Modern Approach},
}

@Article{Robinson196501Machine,
  author     = {Robinson, J. A.},
  journal    = {J. ACM},
  title      = {A Machine-Oriented Logic Based on the Resolution Principle},
  year       = {1965},
  issn       = {0004-5411},
  month      = jan,
  number     = {1},
  pages      = {23–41},
  volume     = {12},
  address    = {New York, NY, USA},
  doi        = {10.1145/321250.321253},
  issue_date = {Jan. 1965},
  numpages   = {19},
  publisher  = {Association for Computing Machinery},
}

@Misc{Sinha2024Wus,
  author        = {Shiven Sinha and Ameya Prabhu and Ponnurangam Kumaraguru and Siddharth Bhat and Matthias Bethge},
  title         = {Wu's Method can Boost Symbolic AI to Rival Silver Medalists and AlphaGeometry to Outperform Gold Medalists at IMO Geometry},
  year          = {2024},
  archiveprefix = {arXiv},
  eprint        = {2404.06405},
}

@InProceedings{Grover1996fast,
author = {Grover, Lov K.},
title = {A fast quantum mechanical algorithm for database search},
year = {1996},
isbn = {0897917855},
url = {https://doi.org/10.1145/237814.237866},
doi = {10.1145/237814.237866},
booktitle = {Proceedings of the Twenty-Eighth Annual ACM Symposium on Theory of Computing},
pages = {212–219},
numpages = {8},
location = {Philadelphia, Pennsylvania, USA},
series = {STOC '96}
}

@Article{Yoder201411Fixed,
  title = {Fixed-Point Quantum Search with an Optimal Number of Queries},
  author = {Yoder, Theodore J. and Low, Guang Hao and Chuang, Isaac L.},
  journal = {Phys. Rev. Lett.},
  volume = {113},
  issue = {21},
  pages = {210501},
  numpages = {5},
  year = {2014},
  month = {Nov},
  publisher = {American Physical Society},
  doi = {10.1103/PhysRevLett.113.210501},
  url = {https://link.aps.org/doi/10.1103/PhysRevLett.113.210501}
}

@Article{Takahashi2009Quantum,
  title={Quantum Arithmetic Circuits: A Survey},
  author={Yasuhiro Takahashi},
  journal={IEICE Transactions on Fundamentals of Electronics, Communications and Computer Sciences},
  volume={E92.A},
  number={5},
  pages={1276-1283},
  year={2009},
  doi={10.1587/transfun.E92.A.1276}
}

@InProceedings{Kabanets2003Derandomizing,
author = {Kabanets, Valentine and Impagliazzo, Russell},
title = {Derandomizing polynomial identity tests means proving circuit lower bounds},
year = {2003},
doi = {10.1145/780542.780595},
booktitle = {Proceedings of the Thirty-Fifth Annual ACM Symposium on Theory of Computing},
pages = {355–364},
numpages = {10},
location = {San Diego, CA, USA},
series = {STOC '03}
}

@Article{Aaronson2014Need,
  author    = {Aaronson, Scott and Ambainis, Andris},
  journal   = {Theory Comput.},
  title     = {The Need for Structure in Quantum Speedups},
  year      = {2014},
  issn      = {1557-2862},
  number    = {1},
  pages     = {133--166},
  volume    = {10},
  doi       = {10.4086/toc.2014.v010a006},
  publisher = {Theory of Computing Exchange},
}

@Article{Buhrman200210Complexity,
  author    = {Buhrman, Harry and de Wolf, Ronald},
  journal   = {Theor. Comput. Sci.},
  title     = {Complexity measures and decision tree complexity: a survey},
  year      = {2002},
  issn      = {0304-3975},
  month     = oct,
  number    = {1},
  pages     = {21--43},
  volume    = {288},
  doi       = {10.1016/s0304-3975(01)00144-x},
  publisher = {Elsevier BV},
}

@Article{Beals200107Quantum,
  author    = {Beals, Robert and Buhrman, Harry and Cleve, Richard and Mosca, Michele and de Wolf, Ronald},
  journal   = {J. ACM},
  title     = {Quantum lower bounds by polynomials},
  year      = {2001},
  issn      = {1557-735X},
  month     = jul,
  number    = {4},
  pages     = {778--797},
  volume    = {48},
  doi       = {10.1145/502090.502097},
  publisher = {Association for Computing Machinery (ACM)},
}

@Article{Bennett199710Strengths,
  author    = {Bennett, Charles H. and Bernstein, Ethan and Brassard, Gilles and Vazirani, Umesh},
  journal   = {SIAM J. Comput.},
  title     = {Strengths and Weaknesses of Quantum Computing},
  year      = {1997},
  issn      = {1095-7111},
  month     = oct,
  number    = {5},
  pages     = {1510--1523},
  volume    = {26},
  doi       = {10.1137/s0097539796300933},
  publisher = {Society for Industrial & Applied Mathematics (SIAM)},
}

@Article{Shende200306Synthesis,
  author    = {Shende, V.V. and Prasad, A.K. and Markov, I.L. and Hayes, J.P.},
  journal   = {IEEE Trans. Comput.-Aided Des. Integr. Circuits Syst.},
  title     = {Synthesis of reversible logic circuits},
  year      = {2003},
  issn      = {0278-0070},
  month     = jun,
  number    = {6},
  pages     = {710--722},
  volume    = {22},
  doi       = {10.1109/tcad.2003.811448},
  publisher = {Institute of Electrical and Electronics Engineers (IEEE)},
}

@Article{Wu202412Asymptotically,
  author    = {Wu, Xian and Li, Lvzhou},
  journal   = {Inf. Comput.},
  title     = {Asymptotically optimal synthesis of reversible circuits},
  year      = {2024},
  issn      = {0890-5401},
  month     = dec,
  pages     = {105235},
  volume    = {301},
  doi       = {10.1016/j.ic.2024.105235},
  publisher = {Elsevier BV},
}

@Article{Deutsch1985Quantum,
  author   = {Deutsch, David and Penrose, Roger},
  journal  = {Proc. R. Soc. Lond. A},
  title    = {Quantum theory, the Church–Turing principle and the universal quantum computer},
  year     = {1985},
  number   = {1818},
  pages    = {97-117},
  volume   = {400},
  doi      = {10.1098/rspa.1985.0070},
  url      = {https://royalsocietypublishing.org/doi/abs/10.1098/rspa.1985.0070},
}

@InProceedings{Bernstein1993Quantum,
  author    = {Bernstein, Ethan and Vazirani, Umesh},
  booktitle = {Proceedings of the Twenty-Fifth Annual ACM Symposium on Theory of Computing},
  title     = {Quantum complexity theory},
  year      = {1993},
  pages     = {11–20},
  series    = {STOC '93},
  doi       = {10.1145/167088.167097},
  isbn      = {0897915917},
  numpages  = {10},
  url       = {https://doi.org/10.1145/167088.167097},
}

@InProceedings{ChiChihYao1993Quantum,
  author     = {Chi-Chih Yao, A.},
  booktitle  = {Proceedings of 1993 IEEE 34th Annual Foundations of Computer Science},
  title      = {Quantum circuit complexity},
  year       = {1993},
  pages      = {352-361},
  series     = {SFCS-93},
  collection = {SFCS-93},
  doi        = {10.1109/SFCS.1993.366852}
}

@article{Shpilka201012Arithmetic,
author = {Shpilka, Amir and Yehudayoff, Amir},
title = {Arithmetic Circuits: A survey of recent results and open questions},
year = {2010},
issue_date = {March 2010},
publisher = {Now Publishers Inc.},
address = {Hanover, MA, USA},
volume = {5},
number = {3–4},
doi = {10.1561/0400000039},
journal = {Found. Trends Theor. Comput. Sci.},
month = mar,
pages = {207–388},
numpages = {182}
}

@book{Loveland2016Automated,
  title={Automated theorem proving: A logical basis},
  author={Loveland, Donald W},
  year={2016},
  publisher={North Holland}
}

@book{Fitting2012First,
  title={First-order logic and automated theorem proving},
  author={Fitting, Melvin},
  year={2013},
  publisher={Springer New York}
}

@Article{Schoenhage197109Schnelle,
  author    = {Schönhage, A. and Strassen, V.},
  journal   = {Computing},
  title     = {Schnelle Multiplikation großer Zahlen},
  year      = {1971},
  issn      = {1436-5057},
  month     = sep,
  number    = {3–4},
  pages     = {281--292},
  volume    = {7},
  doi       = {https://doi.org/10.1007/BF02242355},
  publisher = {Springer Science and Business Media LLC},
}

@Article{Cooley1965algorithm,
  author    = {Cooley, James W. and Tukey, John W.},
  journal   = {Math. Comput.},
  title     = {An algorithm for the machine calculation of complex Fourier series},
  year      = {1965},
  issn      = {1088-6842},
  number    = {90},
  pages     = {297--301},
  volume    = {19},
  doi       = {https://doi.org/10.2307/2003354},
  publisher = {American Mathematical Society (AMS)},
}

@Article{Seidel2022Efficient,
  author    = {Seidel, Raphael and Tcholtchev, Nikolay and Bock, Sebastian and Becker, Colin Kai-Uwe and Hauswirth, Manfred},
  journal   = {IEEE Access},
  title     = {Efficient Floating Point Arithmetic for Quantum Computers},
  year      = {2022},
  issn      = {2169-3536},
  pages     = {72400--72415},
  volume    = {10},
  doi       = {10.1109/ACCESS.2022.3188251},
  publisher = {Institute of Electrical and Electronics Engineers (IEEE)},
}

@article{Wootters1982Single,
  title={A single quantum cannot be cloned},
  author={Wootters, William K and Zurek, Wojciech H},
  journal={Nature},
  volume={299},
  number={5886},
  pages={802--803},
  year={1982},
  publisher={Nature Publishing Group UK London},
url={https://www.nature.com/articles/299802a0},
doi={https://doi.org/10.1038/299802a0}
}

@Article{Schwartz198010Fast,
author = {Schwartz, J. T.},
title = {Fast Probabilistic Algorithms for Verification of Polynomial Identities},
year = {1980},
issue_date = {Oct. 1980},
publisher = {Association for Computing Machinery},
address = {New York, NY, USA},
volume = {27},
number = {4},
issn = {0004-5411},
url = {https://doi.org/10.1145/322217.322225},
doi = {10.1145/322217.322225},
journal = {J. ACM},
month = oct,
pages = {701–717},
numpages = {17}
}

@Article{Hales2017Formal,
  author = {Hales, Thomas and Adams, Mark and Bauer, Gertrud and Dang, Tat Dat and
            Harrison, John and Hoang, Le Truong and Kaliszyk, Cezary and
            Magron, Victor and McLaughlin, Sean and Nguyen, Tat Thang and
            Nguyen, Quang Truong and Nipkow, Tobias and Obua, Steven and
            Pleso, Joseph and Rute, Jason and Solovyev, Alexey and
            Ta, Thi Hoai An and Tran, Nam Trung and Trieu, Thi Diep and
            Urban, Josef and Vu, Ky and Zumkeller, Roland},
  journal   = {Forum Math. Pi},
  title     = {A FORMAL PROOF OF THE KEPLER CONJECTURE},
  year      = {2017},
  issn      = {2050-5086},
  volume    = {5},
  pages={e2},
  doi       = {10.1017/fmp.2017.1},
  publisher = {Cambridge University Press (CUP)},
}

@article{Gonthier2008Formal,
  title={Formal proof--the four-color theorem},
  author={Gonthier, Georges},
  journal={Notices of the AMS},
  volume={55},
  number={11},
  pages={1382--1393},
  year={2008}
}

@article{Guo2025Deepseek,
  title={Deepseek-r1 incentivizes reasoning in llms through reinforcement learning},
    author    = {Guo, Daya and Yang, Dejian and Zhang, Haowei and Song, Junxiao and Wang, Peiyi and Zhu, Qihao and Xu, Runxin and Zhang, Ruoyu and Ma, Shirong and Bi, Xiao and others},
  journal={Nature},
  volume={645},
  number={8081},
  pages={633--638},
  year={2025},
  publisher={Nature Publishing Group UK London},
url={https://doi.org/10.1038/s41586-025-09422-z},
doi={https://doi.org/10.1038/s41586-025-09422-z}
}

@incollection{Hasan2015Formal,
  title={Formal verification methods},
  author={Hasan, Osman and Tahar, Sofiene},
  booktitle={Encyclopedia of Information Science and Technology, Third Edition},
  pages={7162--7170},
  year={2015},
  publisher={IGI Global Scientific Publishing}
}

@InProceedings{Ye2008introduction,
  author    = {Ye, Zheng and Chou, Shang-Ching and Gao, Xiao-Shan},
  booktitle = {Proceedings of the 7th International Conference on Automated Deduction in Geometry},
  title     = {An introduction to java geometry expert},
  year      = {2008},
  pages     = {189–195},
  series    = {ADG'08},
  doi       = {https://doi.org/10.1007/978-3-642-21046-4_10},
  isbn      = {9783642210457},
  issn      = {1611-3349},
  numpages  = {7},
}

@Article{Demillo197806probabilistic,
  author    = {Demillo, Richard A. and Lipton, Richard J.},
  journal   = {Inf. Process. Lett.},
  title     = {A probabilistic remark on algebraic program testing},
  year      = {1978},
  issn      = {0020-0190},
  month     = jun,
  number    = {4},
  pages     = {193--195},
  volume    = {7},
  doi       = {10.1016/0020-0190(78)90067-4},
  publisher = {Elsevier BV},
}

@InProceedings{Zippel1979Probabilistic,
  author    = {Zippel, Richard},
  booktitle = {Symbolic and Algebraic Computation},
  title     = {Probabilistic algorithms for sparse polynomials},
  year      = {1979},
  pages     = {216--226},
  doi       = {10.1007/3-540-09519-5_73},
  isbn      = {978-3-540-35128-3},
}

@Article{Montanaro201601Quantum,
  author    = {Montanaro, Ashley},
  journal   = {npj Quantum Inf.},
  title     = {Quantum algorithms: an overview},
  year      = {2016},
  issn      = {2056-6387},
  month     = jan,
  number    = {1},
  pages     = {15023},
  volume    = {2},
  doi       = {10.1038/npjqi.2015.23},
  publisher = {Springer Science and Business Media LLC},
}

@Book{Arora2009Computational,
  author    = {Arora, Sanjeev},
  publisher = {Cambridge University Press},
  title     = {Computational complexity},
  year      = {2009},
  address   = {Cambridge, UK},
  isbn      = {9780521424264},
  subtitle  = {a modern approach},
}

@book{Chou1994MachineProofs,
  title     = {Machine Proofs in Geometry: Automated Production of Readable Proofs for Geometry Theorems},
  author    = {Chou, Shang-Ching and Gao, Xiao-Shan and Zhang, Jing-Zhong},
  year      = {1994},
  publisher = {World Scientific},
  address   = {Singapore}
}

@Article{Saxena2009Progress,
  author    = {Saxena, Nitin},
  journal   = {Bulletin of the European Association for Theoretical Computer Science},
  title     = {Progress on Polynomial Identity Testing},
  year      = {2009},
  pages     = {49--79},
  volume    = {99},
  publisher = {European Association for Theoretical Computer Science},
}

@InBook{Saxena2014Progress,
  author    = {Saxena, Nitin},
  pages     = {131--146},
  publisher = {Springer International Publishing},
  title     = {Progress on Polynomial Identity Testing-{II}},
  year      = {2014},
  address   = {Cham},
  isbn      = {978-3-319-05446-9},
  booktitle = {Perspectives in Computational Complexity: The Somenath Biswas Anniversary Volume},
  doi       = {10.1007/978-3-319-05446-9_7},
  url       = {https://doi.org/10.1007/978-3-319-05446-9_7},
}

@Article{Borodin197406Fast,
  author    = {Borodin, A. and Moenck, R.},
  journal   = {J. Comput. Syst. Sci.},
  title     = {Fast modular transforms},
  year      = {1974},
  issn      = {0022-0000},
  month     = jun,
  number    = {3},
  pages     = {366--386},
  volume    = {8},
  doi       = {10.1016/s0022-0000(74)80029-2},
  publisher = {Elsevier BV},
}

@Book{Szegoe1939Orthogonal,
  author    = {Szeg{\H{o}}, G{\'a}bor},
  publisher = {American Mathematical Society},
  title     = {Orthogonal Polynomials},
  year      = {1939},
}

@Article{Wu1978Decision,
  author  = {Wen-Tsün Wu},
  journal = {Sci. Sin.},
  title   = {ON THE DECISION PROBLEM AND THE MECHANIZATION OF THEOREM-PROVING IN ELEMENTARY GEOMETRY},
  year    = {1978},
  number  = {2},
  pages   = {159-172},
  volume  = {21},
}

@article{Hubert2025Olympiad,
  title={Olympiad-level formal mathematical reasoning with reinforcement learning},
  author    = {Hubert, Thomas and Mehta, Rishi and Sartran, Laurent and Horváth, Miklós Z. and Žužić, Goran and Wieser, Eric and Huang, Aja and Schrittwieser, Julian and Schroecker, Yannick and Masoom, Hussain and others},
  journal={Nature},
  year={2025},
  publisher={Nature Publishing Group UK London},
doi={https://doi.org/10.1038/s41586-025-09833-y}
}

@article{Jin2025Topological,
  title={Topological prethermal strong zero modes on superconducting processors},
  author    = {Jin, Feitong and Jiang, Si and Zhu, Xuhao and Bao, Zehang and Shen, Fanhao and Wang, Ke and Zhu, Zitian and Xu, Shibo and Song, Zixuan and Chen, Jiachen and others},
  journal={Nature},
  volume={645},
  number={8081},
  pages={626--632},
  year={2025},
  publisher={Nature Publishing Group UK London},
doi={https://doi.org/10.1038/s41586-025-09476-z}
}

@Misc{Wang2025Demonstration,
  author={Ke Wang and Zhide Lu and Chuanyu Zhang and Gongyu Liu and Jiachen Chen and Yanzhe Wang and Yaozu Wu and Shibo Xu and Xuhao Zhu and Feitong Jin and others},
  title         = {Demonstration of low-overhead quantum error correction codes},
  year          = {2025},
  archiveprefix = {arXiv},
  eprint        = {2505.09684},
}

@article{Gupta2024Encoding,
  title = {Encoding a Magic State with beyond Break-Even Fidelity},
  year = {2024},
  journal = {Nature},
  volume = {625},
  number = {7994},
  pages = {259--263},
  issn = {0028-0836, 1476-4687},
  doi = {10.1038/s41586-023-06846-3},
  author    = {Gupta, Riddhi S. and Sundaresan, Neereja and Alexander, Thomas and Wood, Christopher J. and Merkel, Seth T. and Healy, Michael B. and Hillenbrand, Marius and Jochym-O’Connor, Tomas and Wootton, James R. and Yoder, Theodore J. and others},
}

@Article{Acharya2025,
author={{Google Quantum AI} and others},
title={Quantum error correction below the surface code threshold},
journal={Nature},
year={2025},
month={Feb},
day={01},
volume={638},
number={8052},
pages={920-926},
issn={1476-4687},
doi={10.1038/s41586-024-08449-y},
url={https://doi.org/10.1038/s41586-024-08449-y}
}

@article{Biamonte2017Quantum,
  title = {Quantum Machine Learning},
  author = {Biamonte, Jacob and Wittek, Peter and Pancotti, Nicola and Rebentrost, Patrick and Wiebe, Nathan and Lloyd, Seth},
  year = {2017},
  month = sep,
  journal = {Nature},
  volume = {549},
  number = {7671},
  pages = {195--202},
  publisher = {Nature Publishing Group},
  doi = {10.1038/nature23474}
}

@article{Dunjko2018Machine,
  title = {Machine Learning \& Artificial Intelligence in the Quantum Domain: A Review of Recent Progress},
  shorttitle = {Machine Learning \& Artificial Intelligence in the Quantum Domain},
  author = {Dunjko, Vedran and Briegel, Hans J.},
  year = {2018},
  month = jun,
  journal = {Rep. Prog. Phys.},
  volume = {81},
  number = {7},
  pages = {074001},
  publisher = {IOP Publishing},
  doi = {10.1088/1361-6633/aab406}
}

@article{DasSarma2019Machine,
  title = {Machine Learning Meets Quantum Physics},
  author = {Das Sarma, Sankar and Deng, Dong-Ling and Duan, Lu-Ming},
  year = {2019},
  month = mar,
  journal = {Phys. Today},
  volume = {72},
  number = {3},
  pages = {48--54},
  publisher = {American Institute of Physics},
  doi = {10.1063/PT.3.4164}
}

@article{Cerezo2022Challenges,
  title = {Challenges and Opportunities in Quantum Machine Learning},
  author = {Cerezo, M. and Verdon, Guillaume and Huang, Hsin-Yuan and Cincio, Lukasz and Coles, Patrick J.},
  year = {2022},
  month = sep,
  journal = {Nat. Comput. Sci.},
  volume = {2},
  number = {9},
  pages = {567--576},
  publisher = {Nature Publishing Group},
  doi = {10.1038/s43588-022-00311-3}
}

@article{Li2025Pitfalls,
  title={Pitfalls and prospects of quantum machine learning},
  author={Li, Weikang and Ma, Yixuan and Deng, Dong-Ling},
  journal={Nat. Comput. Sci.},
volume={5},
  pages={1095–1097},
  year={2025},
  publisher={Nature Publishing Group},
url={https://doi.org/10.1038/s43588-025-00914-6},
doi={https://doi.org/10.1038/s43588-025-00914-6}
}

@article{Harrow2009Quantum,
  title = {Quantum {{Algorithm}} for {{Linear Systems}} of {{Equations}}},
  author = {Harrow, Aram W. and Hassidim, Avinatan and Lloyd, Seth},
  year = {2009},
  month = oct,
  journal = {Phys. Rev. Lett.},
  volume = {103},
  number = {15},
  pages = {150502},
  publisher = {American Physical Society},
  doi = {10.1103/PhysRevLett.103.150502}
}

@article{Havlicek2019Supervised,
  title = {Supervised Learning with Quantum-Enhanced Feature Spaces},
  author = {Havl{\'i}{\v c}ek, Vojt{\v e}ch and C{\'o}rcoles, Antonio D. and Temme, Kristan and Harrow, Aram W. and Kandala, Abhinav and Chow, Jerry M. and Gambetta, Jay M.},
  year = {2019},
  month = mar,
  journal = {Nature},
  volume = {567},
  number = {7747},
  pages = {209--212},
  publisher = {Nature Publishing Group},
  doi = {10.1038/s41586-019-0980-2}
}

@article{Lloyd2014Quantum,
  title = {Quantum Principal Component Analysis},
  author = {Lloyd, Seth and Mohseni, Masoud and Rebentrost, Patrick},
  year = {2014},
  month = sep,
  journal = {Nat. Phys.},
  volume = {10},
  number = {9},
  pages = {631--633},
  publisher = {Nature Publishing Group},
  doi = {10.1038/nphys3029}
}

@article{Gao2018Quantum,
  title = {A Quantum Machine Learning Algorithm Based on Generative Models},
  author = {Gao, X. and Zhang, Z.-Y. and Duan, L.-M.},
  year = {2018},
  month = dec,
  journal = {Sci. Adv.},
  volume = {4},
  number = {12},
  pages = {eaat9004},
  publisher = {American Association for the Advancement of Science},
  doi = {10.1126/sciadv.aat9004},
  chapter = {Research Article}
}

@article{Liu2021Rigorous,
  title = {A Rigorous and Robust Quantum Speed-up in Supervised Machine Learning},
  author = {Liu, Yunchao and Arunachalam, Srinivasan and Temme, Kristan},
  year = {2021},
  month = sep,
  journal = {Nat. Phys.},
  volume = {17},
  number = {9},
  pages = {1013--1017},
  publisher = {Nature Publishing Group},
  doi = {10.1038/s41567-021-01287-z}
}

@article{Hu2019Quantum,
  title = {Quantum Generative Adversarial Learning in a Superconducting Quantum Circuit},
  author = {Hu, Ling and Wu, Shu-Hao and Cai, Weizhou and Ma, Yuwei and Mu, Xianghao and Xu, Yuan and Wang, Haiyan and Song, Yipu and Deng, Dong-Ling and Zou, Chang-Ling and Sun, Luyan},
  year = {2019},
  month = jan,
  journal = {Sci. Adv.},
  volume = {5},
  number = {1},
  pages = {eaav2761},
  publisher = {American Association for the Advancement of Science},
  doi = {10.1126/sciadv.aav2761},
  chapter = {Research Article}
}

@article{Ren2022Experimental,
  title = {Experimental Quantum Adversarial Learning with Programmable Superconducting Qubits},
  author = {Ren, Wenhui and Li, Weikang and Xu, Shibo and Wang, Ke and Jiang, Wenjie and Jin, Feitong and Zhu, Xuhao and Chen, Jiachen and Song, Zixuan and Zhang, Pengfei and others},
  year = {2022},
  month = nov,
  journal = {Nat. Comput. Sci.},
  volume = {2},
  number = {11},
  pages = {711--717},
  publisher = {Nature Publishing Group},
  doi = {10.1038/s43588-022-00351-9}
}

@article{Jerbi2023Quantum,
  title = {Quantum Machine Learning beyond Kernel Methods},
  author = {Jerbi, Sofiene and Fiderer, Lukas J. and Poulsen Nautrup, Hendrik and K{\"u}bler, Jonas M. and Briegel, Hans J. and Dunjko, Vedran},
  year = {2023},
  month = jan,
  journal = {Nat. Commun.},
  volume = {14},
  number = {1},
  pages = {517},
  publisher = {Nature Publishing Group},
  doi = {10.1038/s41467-023-36159-y}
}

@article{Dunjko2016Quantum,
  title = {Quantum-Enhanced Machine Learning},
  author = {Dunjko, Vedran and Taylor, Jacob M. and Briegel, Hans J.},
  journal = {Phys. Rev. Lett.},
  volume = {117},
  issue = {13},
  pages = {130501},
  numpages = {6},
  year = {2016},
  month = {Sep},
  publisher = {American Physical Society},
  doi = {10.1103/PhysRevLett.117.130501},
  url = {https://link.aps.org/doi/10.1103/PhysRevLett.117.130501}
}

@article{Lacroix2025Scaling,
  title={Scaling and logic in the color code on a superconducting quantum processor},
  author={Lacroix, Nathan and Bourassa, Alexandre and Heras, Francisco JH and Zhang, Lei M and Bausch, Johannes and Senior, Andrew W and Edlich, Thomas and Shutty, Noah and Sivak, Volodymyr and Bengtsson, Andreas and others},
  journal={Nature},
volume={645},
  pages={614},
  year={2025},
  publisher={Nature Publishing Group UK London},
doi={https://doi.org/10.1038/s41586-025-09061-4}
}

@article{Huang2022Quantum,
  title={Quantum advantage in learning from experiments},
  author={Huang, Hsin-Yuan and Broughton, Michael and Cotler, Jordan and Chen, Sitan and Li, Jerry and Mohseni, Masoud and Neven, Hartmut and Babbush, Ryan and Kueng, Richard and Preskill, John and others},
  journal={Science},
  volume={376},
  number={6598},
  pages={1182--1186},
  year={2022},
  publisher={American Association for the Advancement of Science},
url={https://www.science.org/doi/10.1126/science.abn7293},
doi={10.1126/science.abn7293}
}

@misc{IMOOfficialWebsite,
  title        = {International Mathematical Olympiad},
  howpublished = {\url{https://www.imo-official.org}},
  year         = {2025},
  note         = {Accessed: 2026-01-12}
}

\let\addcontentsline\oldaddcontentsline

\clearpage
\newpage 
\onecolumngrid
\setcounter{section}{0}
\setcounter{equation}{0}
\setcounter{figure}{0}
\setcounter{table}{0}
\makeatletter
\renewcommand\thefigure{S\arabic{figure}}
\renewcommand\thetable{S\arabic{table}}
\renewcommand\theequation{S\arabic{equation}}

\begin{center} 
	{\large \bf Supplementary Information: Quantum automated theorem proving}
\end{center} 

\maketitle
\tableofcontents

\section{Preliminaries of automated theorem proving and propositional logic}

We begin by reviewing the standard framework of automated theorem proving (ATP) in propositional logic \cite{Loveland2016Automated,Russell2020Artificial}. As a logic-based reasoning paradigm, ATP represents knowledge using formal sentences and derives conclusions by applying well-defined inference rules. Its defining characteristic is a deterministic, rule-based reasoning process, in which a knowledge base (KB) is encoded as a collection of formulas and logical consequences are derived through systematic inference.

A knowledge base is a set of sentences expressed according to the syntax of a given logical representation. For example, in elementary arithmetic, the expression ``$x+y=4$'' is a well-formed sentence, whereas ``$x4y+=$'' is not. Beyond syntax, a logical system also requires a definition of semantics, which assigns meaning to sentences by specifying their truth values in different possible worlds. In arithmetic, the sentence ``$x+y=4$'' is true in a world where $x=2$ and $y=2$, but false in a world where $x=1$ and $y=1$. A model is a mathematical abstraction representing such a possible world. If a sentence $\alpha$ is true in a model $m$, we say that $m$ satisfies $\alpha$, or that $m$ is a model of $\alpha$. We denote by $M(\alpha)$ the set of all models of $\alpha$.

Logical reasoning is formalized through the notion of logical entailment, written as
\begin{align}\label{eq-entail}
    \alpha \models \beta ,
\end{align}
which means that the sentence $\beta$ is a logical consequence of the sentence $\alpha$. Formally, $\alpha \models \beta$ if and only if
\begin{align}
M(\alpha) \subseteq M(\beta).
\end{align}

In automated reasoning, new sentences are repeatedly derived from those already present in the knowledge base by applying inference rules. This process continues until no further derivations are possible or until a contradiction is obtained. An inference procedure is said to be sound (or truth-preserving) if every sentence it derives is logically entailed by the knowledge base. Soundness guarantees that no false statements are produced. The complementary notion is completeness: an inference procedure is complete if it can derive every sentence that is logically entailed by the knowledge base. Together, soundness and completeness ensure that all and only logically valid consequences can be derived~\cite{Russell2020Artificial}.

Two inference rules are particularly prominent in propositional logic: modus ponens and resolution. Modus ponens--based reasoning is sound and complete for Horn clauses and can be implemented in linear time. However, Horn clauses are restricted to disjunctions containing at most one positive literal. In this work, we focus on resolution, which applies to arbitrary knowledge bases in propositional logic and extends naturally to first-order logic (FOL). Resolution is typically used in a refutation-based manner and constitutes a fundamental inference mechanism in classical ATP.

The resolution rule in propositional logic takes the form
\begin{align}\label{eq-resolution-propostional}
    {{a_1 \vee \cdots \vee a_n \vee c,\quad b_1 \vee \cdots \vee b_m \vee \neg c}
    \over
    {a_1 \vee \cdots \vee a_n \vee b_1 \vee \cdots \vee b_m}},
\end{align}
where all $a_i$, $b_i$, and $c$ are literals, and the horizontal line denotes logical entailment. A literal is either a propositional variable or its negation. The logical connectives $\vee$, $\wedge$, and $\neg$ denote disjunction, conjunction, and negation, respectively. In this manuscript, we occasionally use a comma to denote conjunction as an alternative to $\wedge$.

Several examples of resolution are given in Eq.~(\ref{eq-example-resolution}):

	\begin{subequations}\label{eq-example-resolution}
		\begin{align}\label{eq-resolution-normal}
			{{A \vee C,B \vee \neg C} \over {A \vee B}},
		\end{align}
		\begin{align}\label{eq-resolution-same}
			{{A \vee C \vee D,B \vee \neg C \vee D} \over {A \vee B \vee D}},
		\end{align}
		\begin{align}\label{eq-resolution-fail}
			{{A \vee C \vee D,B \vee \neg C \vee \neg D} \over {A \vee C \vee D,B \vee \neg C \vee \neg D}},
		\end{align}
		\begin{align}\label{eq-resolution-empty}
			{{C,\neg C} \over {\bot}}
		\end{align}
	\end{subequations}
Here, $A$, $B$, $C$, and $D$ are propositional variables. Equation~(\ref{eq-resolution-normal}) illustrates a typical resolution step. In Eq.~(\ref{eq-resolution-same}), both premises contain the same literal $D$, which appears only once in the conclusion; this simplification is known as factoring. Equation~(\ref{eq-resolution-fail}) shows that when more than one pair of complementary literals appears across the premises, no resolution is possible. Finally, Eq.~(\ref{eq-resolution-empty}) produces the empty clause $\bot$, which represents a contradiction~\cite{Russell2020Artificial}. The derivation of an empty clause signals inconsistency of the knowledge base and terminates the ATP process.

Resolution applies to formulas expressed in conjunctive normal form (CNF), defined as a conjunction of clauses, each of which is a disjunction of literals. Every propositional logic formula is logically equivalent to a CNF formula. A standard transformation procedure proceeds as follows. We begin with formulas composed of the connectives $\neg$, $\vee$, and $\wedge$, propositional variables, and parentheses. Any additional connectives (such as implication or biconditional) are first eliminated using logical equivalences. Negations are then pushed inward using De Morgan’s laws:
\begin{subequations}\label{eq-move-negation-parenthese}
\begin{align}
    {\neg (a \wedge b)} &\equiv {\neg a \vee \neg b}, \\
    {\neg (a \vee b)} &\equiv {\neg a \wedge \neg b}. \label{eq-move-negation-disjunction}
\end{align}
\end{subequations}
Double negations are removed using
\begin{align}\label{eq-eliminate-double-negation}
    {\neg \neg a} \equiv a,
\end{align}
and finally, disjunction is distributed over conjunction:
\begin{align}\label{eq-distribute-disjunction}
    {(a \wedge b) \vee c} \equiv {(a \vee c) \wedge (b \vee c)}.
\end{align}
Repeated application of these rules yields a conjunction of clauses, which constitutes the CNF representation of the knowledge base.

To establish an entailment of the form $\text{KB} \models \alpha$, resolution-based inference proceeds via proof by contradiction. One constructs the conjunction $\text{KB} \wedge \neg \alpha$, converts it into CNF, and applies resolution iteratively. At each step, pairs of clauses containing complementary literals are resolved to produce new clauses, which are added to the set if not already present. This process continues until either no further clauses can be generated—indicating that KB does not entail $\alpha$—or the empty clause $\bot$ is derived, which certifies that $\text{KB} \models \alpha$.

\section{Quantum automated theorem proving in propositional logic and first-order logic}

\subsection{Quantum resolution in propositional logic}
\label{section:quantum resolution in propositional logic}

In quantum automated theorem proving (QATP), the knowledge base—such as a collection of clauses in propositional logic—is encoded and processed directly on a quantum computer. We employ a quantum circuit to load the knowledge base into a quantum state. Concretely, this encoding is implemented by a unitary operator that takes as input a quantum register encoding the indices of clauses and entangles it with a data register storing the corresponding clause information:
\begin{align}\label{eq-qram-u}
		{\left| j \right\rangle _\text{I}}{\left| 0 \right\rangle _\text{D}} \to {\left| j \right\rangle _\text{I}}{\left| {{m_j}} \right\rangle _\text{D}},
	\end{align}
where ${\left| j \right\rangle _\text{I}}$ denotes the state of the index register and ${\left| {{m_j}} \right\rangle _\text{D}}$ encodes the corresponding clause or coefficient in the data register.

To perform reasoning in propositional logic, it is necessary to implement the resolution rule on a quantum computer. A central challenge arises from the fact that classical resolution is inherently irreversible: the premises cannot, in general, be reconstructed from the conclusion. In contrast, quantum circuits must be unitary and therefore reversible.
In QATP, we address this challenge by introducing auxiliary qubits that store the resolution outcome while preserving the quantum states representing the premises. The basic quantum resolution operation is defined as
	\begin{align}\label{eq-qresolution}
		{U_\text{R}}\left| {{c_1}} \right\rangle \left| {{c_2}} \right\rangle \left| 0 \right\rangle  = \left| {{c_1}} \right\rangle \left| {{c_2}} \right\rangle \left| \mathcal{R}(c_1, c_2) \right\rangle,
	\end{align}
where $\left| {{c_1}} \right\rangle $ and $\left| {{c_2}} \right\rangle $ encode the premise clauses and $|\mathcal{R}(c_1,c_2)\rangle$ represents the resolution result. The process of quantum
resolution is similar to the classical counterpart shown in Eq.~(\ref{eq-resolution-propostional}). This construction ensures reversibility by retaining the original premises.

The explicit form of the quantum resolution circuit depends on the chosen clause representation. In CNF representation, each propositional variable is associated with a fixed register position, and its presence and polarity are encoded in the local quantum state. Logical disjunction symbols are omitted, as clauses are represented implicitly by the joint state of all variables. For example, the clause $A \vee \neg C$ can be encoded as
${\left| { 1} \right\rangle _A}{\left| 0 \right\rangle _B}{\left| {2} \right\rangle _C}{\left| 0 \right\rangle _D}$,
where the four registers correspond to the propositional variables $\{A,B,C,D\}$. The state $|1\rangle$ denotes the presence of a positive literal, $|2\rangle$ denotes the presence of a negated literal, and $|0\rangle$ denotes absence. The four basis states $|0\rangle$, $|1\rangle$, $|2\rangle$, and $|3\rangle$ form a ququart (four-level system), which can equivalently be represented using two qubits as $|00\rangle$, $|01\rangle$, $|10\rangle$, and $|11\rangle$.

\begin{figure}[htb]
\centering
\includegraphics[width=0.25\linewidth]{\figpath/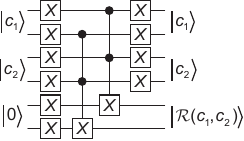}
\caption{\label{fig-quantum-resolution-circuit}Quantum circuit implementing the resolution operation for a single propositional variable.}
\end{figure}

Quantum resolution circuit operates by examining propositional variables one by one.
For each variable, its occurrences in the two premise clauses determine how it is handled. In particular, a variable may appear in the two premises with opposite polarities, with the same polarity, in only one premise, or in neither premise, as shown in Eq.~(\ref{eq-example-resolution}). We first describe the unitary implementation for the representative situation shown in Eq.~(\ref{eq-resolution-normal}), and then explain how a verification circuit is used to judge the validity of all resolution situations shown in Eq. (\ref{eq-example-resolution}).
In Eq.~(\ref{eq-resolution-normal}), exactly one propositional variable (e.g., $C$) appears in the two premises with opposite polarities and is eliminated in the conclusion. The remaining propositional variables occur only in one clause of the premise and are recorded in the conclusion. The corresponding unitary action on a single propositional variable is
\begin{subequations}\label{eq-quantum-resolution-qram}
		\begin{align}
			{U_\text{R}}\left| {0} \right\rangle \left| {0} \right\rangle \left| {0} \right\rangle  = \left| {0} \right\rangle \left| {0} \right\rangle \left| {0} \right\rangle ,
		\end{align}
		\begin{align}
			{U_\text{R}}\left| {0} \right\rangle \left| {1} \right\rangle \left| {0} \right\rangle  = \left| {0} \right\rangle \left| {1} \right\rangle \left| {1} \right\rangle ,
		\end{align}
		\begin{align}
			{U_\text{R}}\left| {0} \right\rangle \left| {2} \right\rangle \left| {0} \right\rangle  = \left| {0} \right\rangle \left| 	{2} \right\rangle \left| {2} \right\rangle ,
		\end{align}
		\begin{align}
			{U_\text{R}}\left| {1} \right\rangle \left| {0} \right\rangle \left| {0} \right\rangle  = \left| {1} \right\rangle \left| 	{0} \right\rangle \left| {1} \right\rangle ,
		\end{align}
		\begin{align}
			{U_\text{R}}\left| {1} \right\rangle \left| {1} \right\rangle \left| {0} \right\rangle  = \left| {1} \right\rangle \left| 	{1} \right\rangle \left| {1} \right\rangle ,
		\end{align}
		\begin{align}
			{U_\text{R}}\left| {1} \right\rangle \left| {2} \right\rangle \left| {0} \right\rangle  = \left| {1} \right\rangle \left| 	{2} \right\rangle \left| {3} \right\rangle ,
		\end{align}
		\begin{align}
			{U_\text{R}}\left| {2} \right\rangle \left| {0} \right\rangle \left| {0} \right\rangle  = \left| {2} \right\rangle \left| 	{0} \right\rangle \left| {2} \right\rangle ,
		\end{align}
		\begin{align}
			{U_\text{R}}\left| {2} \right\rangle \left| {1} \right\rangle \left| {0} \right\rangle  = \left| {2} \right\rangle \left| 	{1} \right\rangle \left| {3} \right\rangle ,
		\end{align}
		\begin{align}
			{U_\text{R}}\left| {2} \right\rangle \left| {2} \right\rangle \left| {0} \right\rangle  = \left| {2} \right\rangle \left| 	{2} \right\rangle \left| {2} \right\rangle ,
		\end{align}
	\end{subequations}
Here, the first two ququarts encode the premise literals and the third ququart encodes the conclusion. The four states $|0\rangle$, $|1\rangle$, $|2\rangle$, and $|3\rangle$ represent absence, positive literal, negative literal, and resolved status, respectively.
For a knowledge base with $N$ propositional variables, $N$ such resolution circuits are applied in parallel—one for each variable. When each ququart is encoded using two qubits, the full quantum resolution circuit takes the form shown in Fig.~\ref{fig-quantum-resolution-circuit}.

Given a knowledge base containing $M$ clauses, the resolution circuit acts on the tensor product of two uniform superpositions over these clauses, producing $M^2$ candidate resolvents. Only resolvents in which exactly one propositional variable is marked as resolved (state $|3\rangle$) are considered valid; all other cases correspond to invalid resolutions, such as the situation shown in Eqs.~(\ref{eq-resolution-fail}), and are discarded. Valid resolvents are added to the knowledge base for the next resolution round.

\begin{figure}[htb]
\centering
\includegraphics[width=1\linewidth]{\figpath/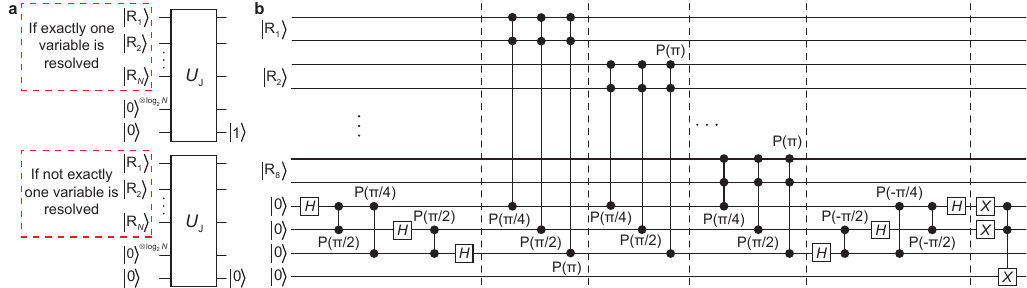}
\caption{\label{fig-judge-sucess-resolution-circuit}Quantum circuit $U_\text{J}$ for validating resolution outcomes. \textbf{a}, Schematic illustration of $U_\text{J}$, which outputs $\left| 1 \right\rangle $ if and only if exactly one propositional variable is in the state $\left| 3 \right\rangle $. \textbf{b}, Explicit construction for $N=8$. A quantum Fourier transform–based addition circuit counts the number of resolved variables, and a controlled Pauli-$X$ gate flags valid resolvents. We first apply the quantum Fourier transform to the $\log_2 N$ auxiliary qubits, preparing the uniform superposition $\sum_{k=0}^{7} \lvert k \rangle / (2\sqrt{2})$. For each resolution result $R_n$ in the state $\lvert 11 \rangle$, a phase factor $\exp(2\pi i k / 8)$ is added to the state on auxiliary qubits. After applying the inverse quantum Fourier transform, the auxiliary register encodes the total number of resolution results in the state $\lvert 3 \rangle$. Finally, a Pauli-$X$ gate controlled on the auxiliary register being in the state $\lvert 001 \rangle$ is used to output states indicating the validity of the result sentence onto the last qubit.}
\end{figure}

To determine whether a candidate resolvent is valid, we introduce a unitary circuit $U_\text{J}$ that checks whether exactly one propositional variable is in the state $|3\rangle$. This circuit takes $N$ resolution outputs as input, uses $\log_2 N$ auxiliary qubits, and writes the result to a single output qubit. The output is $|1\rangle$ if the resolvent is valid and $|0\rangle$ otherwise, as illustrated in Fig.~\ref{fig-judge-sucess-resolution-circuit}\textbf{a}. An explicit construction for $N=8$ is shown in Fig.~\ref{fig-judge-sucess-resolution-circuit}\textbf{b}. In this construction,  we apply a quantum Fourier transform–based addition circuit to the auxiliary qubits, controlled by the resolution results. More specifically, this circuit adds one to the binary value stored in the auxiliary qubits whenever a resolution result $R_n$ is in the state $\lvert 3 \rangle$ (corresponding to $\lvert 11 \rangle$ in the qubit representation). In the final step, a Pauli-$X$ gate controlled on the auxiliary qubits being in the state $\lvert 001 \rangle$ is applied, producing the output state $\lvert 1 \rangle$ if and only if a valid conclusion is reached, that is, exactly one propositional variable is in the state $\lvert 3 \rangle$. The depth of this circuit is $O(N \log N)$.

As a concrete example, consider the knowledge base $\{A \vee \neg C,\; B \vee C,\; \neg B\}$. This knowledge base is encoded as a uniform superposition of the quantum states ${\left| { 1} \right\rangle _A}{\left| 0 \right\rangle _B}{\left| { 2} \right\rangle _C}$, ${\left| {0} \right\rangle _A}{\left| 1 \right\rangle _B}{\left| { 1} \right\rangle _C}$, and ${\left| { 0} \right\rangle _A}{\left| 2\right\rangle _B}{\left| {0} \right\rangle _C}$.
The direct product of two such superpositions forms the input to the quantum resolution circuit. Measurement of the output register in the computational (Pauli-$Z$) basis yields candidate resolvents. The termination conditions of quantum resolution mirror those of the classical procedure: the algorithm halts when either no new clauses can be derived or an empty clause—represented by the all-zero state—is produced.

\subsection{Quantum resolution in first-order logic}

First-order logic \cite{Fitting2012First}, also known as predicate logic, extends propositional logic by allowing statements about objects and their relations. A predicate denotes a Boolean-valued relation on a fixed number of arguments, returning either \emph{True} or \emph{False}. First-order logic also permits function symbols that map objects to objects, as well as constant symbols denoting specific (but model-dependent) elements of the domain. Unlike propositional variables, which range only over truth values, variables in first-order logic range over objects in the domain of a model. For example, the natural-language statement ``John’s cat is a vehicle'' can be expressed as $\mathrm{IsVehicle}(\mathrm{Cat}(\mathrm{John}))$, where $\mathrm{IsVehicle}$ is a predicate symbol, $\mathrm{Cat}$ is a function symbol, and $\mathrm{John}$ is a constant symbol whose interpretation may vary across models.

First-order logic additionally includes quantifiers that express statements about collections of objects. Universal quantification $\forall$ asserts that a property holds for \emph{all} objects, whereas existential quantification $\exists$ asserts that there exists \emph{some} object for which a property holds. For instance, ``All cats are vehicles'' can be written as $\forall x\bigl(\mathrm{IsCat}(x)\rightarrow \mathrm{IsVehicle}(x)\bigr)$, and ``Some cats are vehicles'' as $\exists x\bigl(\mathrm{IsCat}(x)\wedge \mathrm{IsVehicle}(x)\bigr)$.
In this work, we focus on first-order logic without equality, in which any sentence can be rewritten as a set of clauses in CNF~\cite{Russell2020Artificial}.

\paragraph{Conversion to CNF.}
To apply resolution in first-order logic, sentences must first be converted to CNF. As in propositional logic, negations are pushed inward; in addition, negated quantifiers are transformed using
\begin{subequations}
\begin{align}\label{eq-move-negation-universal-quantifier}
    \neg \forall x\, P \;\Rightarrow\; \exists x\, \neg P,\\
    \neg \exists x\, P \;\Rightarrow\; \forall x\, \neg P,
\end{align}
\end{subequations}
where the symbol $\Rightarrow$ indicates a syntactic rewrite rather than logical implication. Double negations are eliminated using Eq.~(\ref{eq-eliminate-double-negation}). Next, bound variables are \emph{standardized apart}: bound variables are renamed so that no variable symbol appears in more than one quantifier scope, preventing unintended interactions when quantifiers are later removed.

Existential quantifiers are then eliminated by \emph{Skolemization}. In the simplest case, $\exists x\, P(x)$ can be replaced by $P(A)$, where $A$ is a fresh constant symbol. More generally, if the existential variable is within the scope of universal quantifiers, it is replaced by a new \emph{Skolem function} of the universally quantified variables. As an example, consider the sentence “Either there is some sport that x likes, or x has doctors”, which in the first-order logic reads:
\begin{align}\label{eq-cnf-example-fol}
\forall x\bigl[\exists y\bigl(\mathrm{Sport}(y)\wedge \mathrm{Likes}(x,y)\bigr)\bigr]\;\vee\;\bigl[\exists z\,\mathrm{Doctor}(x,z)\bigr].
\end{align}
Skolemization yields
\begin{align}
\forall x\bigl(\mathrm{Sport}(F(x))\wedge \mathrm{Likes}(x,F(x))\bigr)\;\vee\;\mathrm{Doctor}(x,G(x)),
\end{align}
where $F$ and $G$ are Skolem functions. After Skolemization, the existential content is captured by these functions, and the remaining universal quantifiers may be dropped without loss of meaning, since all remaining variables are implicitly universally quantified. Finally, distributing $\vee$ over $\wedge$ using Eq.~(\ref{eq-distribute-disjunction}) produces the CNF:
\begin{align}
&\bigl[\mathrm{Sport}(F(x)) \vee \mathrm{Doctor}(x,G(x))\bigr] \wedge \nonumber\\
&\bigl[\mathrm{Likes}(x,F(x)) \vee \mathrm{Doctor}(x,G(x))\bigr].
\end{align}

\paragraph{Resolution and unification.}
Resolution in first-order logic can be viewed as a lifted generalization of propositional resolution Eq.~(\ref{eq-resolution-propostional}). As in propositional logic, two clauses may be resolved when they contain complementary literals; in first-order logic, two literals are complementary if one can be made identical to the negation of the other by a suitable \emph{substitution}. Determining such a substitution is the task of \emph{unification}, which finds a mapping from variables to terms that makes two expressions syntactically identical. For example,
\begin{align}
\mathrm{UNIFY}\bigl(\mathrm{Doctor}(\mathrm{John},x),\mathrm{Doctor}(\mathrm{John},\mathrm{Jane})\bigr)=\{x/\mathrm{Jane}\}.
\end{align}
Here, $\{x/\mathrm{Jane}\}$ replaces the variable $x$ with the term $\mathrm{Jane}$. We write $\mathrm{SUBST}(\theta,p)$ for the result of applying substitution $\theta$ to an expression $p$. Unification can then be defined abstractly as
\begin{align}
\mathrm{UNIFY}(p,q)=\theta \quad \text{where}\quad \mathrm{SUBST}(\theta,p)=\mathrm{SUBST}(\theta,q).
\end{align}
Unlike an \emph{interpretation}, which maps variables to domain objects, a substitution operates purely at the syntactic level by replacing variables with terms.

With this notation, the (binary) resolution rule in first-order logic is given by
\begin{align}\label{eq-resolution-fol}
{{a_1 \vee \cdots \vee a_n \vee c_1,\quad b_1 \vee \cdots \vee b_m \vee c_2}
\over
{\mathrm{SUBST}\!\left(\theta,\, a_1 \vee \cdots \vee a_n \vee b_1 \vee \cdots \vee b_m\right)}},
\end{align}
where $\mathrm{UNIFY}(c_1,\neg c_2)=\theta$.

For instance, the clauses $\mathrm{Sport}(F(x)) \vee \mathrm{Likes}(G(x),x)$ and $\neg \mathrm{Likes}(u,v)\vee \neg \mathrm{Doctor}(u,v)$ can be resolved by eliminating the complementary literals $\mathrm{Likes}(G(x),x)$ and $\neg \mathrm{Likes}(u,v)$ using the unifier $\theta=\{u/G(x),\, v/x\}$, yielding the resolvent
\begin{align}
\mathrm{Sport}(F(x)) \vee \neg \mathrm{Doctor}(G(x),x).
\end{align}
This rule is called \emph{binary resolution} because it resolves exactly one complementary pair of literals from two clauses.

In first-order logic, binary resolution alone is not complete. Completeness is obtained by combining binary resolution with \emph{factoring}, which merges two unifiable literals within a single clause and applies the corresponding unifier to the entire clause. The combination of binary resolution and factoring yields a complete refutation procedure for first-order logic without equality~\cite{Russell2020Artificial}.

	Now we give a concrete example on the badminton-basketball reasoning task mentioned in the main text (see Fig. 1\textbf{a}). We start with sentences in English to form the knowledge base and state the conclusion we want to deduce:
	\begin{enumerate}
            \item Anyone likes sports that their father likes.
            \item If someone likes badminton, then this person likes a certain type of badminton racket.
            \item Alice does not like any kind of badminton racket.
            \item Badminton and basketball are sports.
            \item People who like the same sports play together.
            \item People who play together either both like badminton or both like basketball.
            \item Alice and Bob play together.
            \item Alice's father and Bob's father do not play together.
            \item Alice plays with her father.
            \item Bob plays with his father.
            \item Bob likes basketball and badminton.
	\end{enumerate}
	We assume that the first ten sentences belong to the knowledge base, and the last sentence is the conclusion we want to infer. We express these sentences in first-order logic as:
	\begin{enumerate}
		\item $ \forall x, y \, \text{Likes}[\text{Father}(x),y] \wedge \text{Sport}(y) \to \text{Likes}(x,y) $;
            \item $ \forall x \, \text{Likes}(x, \text{badminton}) \to [\exists y \, \text{Badmintonracket}(y) \wedge \text{Likes}(x, y)] $
            \item $\forall x \,  \text{Badmintonracket}(x) \to \neg \text{Likes(Alice, x)}$;
            \item $\text{Sport(badminton)}$;
            \item $\text{Sport(basketball)}$;
            \item $\forall x, y, z \, \text{Sport}(x) \wedge \text{Likes}(y,x) \wedge \text{Likes}(z,x) \to \text{Playtogether}(y,z)$;
            \item $\forall x, y \, \text{Playtogether}(x,y) \to [(\text{Likes}(x, \text{badminton}) \wedge \text{Likes}(y, \text{badminton})) \vee (\text{Likes}(x, \text{basketball}) \wedge \text{Likes}(y, \text{basketball}))] $;
            \item $ \text{Playtogether[Alice, Bob]}$;
            \item $ \neg \text{Playtogether[Father(Alice), Father(Bob)]}$;
            \item $ \text{Playtogether[Alice, Father(Alice)]}$;
            \item $ \text{Playtogether[Bob, Father(Bob)]}$;
            \item $ \text{Likes(Bob, basketball)} \wedge \text{Likes(Bob, badminton)}$.
 	\end{enumerate}
	The first eleven sentences are in the knowledge base and the last sentence is the conclusion we want to infer. Recalling that we prove $\text{KB} \models \alpha $ by proving the unsatisfiability of the conjunction $\text{KB} \wedge \neg \alpha $. We transform these sentences to CNF as:
	\begin{enumerate}
		  \item $ \neg \text{Likes[Father}(x),y] \vee \neg \text{Sport}(y) \vee \text{Likes}(x,y)$;
		  \item $ \neg \text{Likes}(x, \text{badminton}) \vee \text{Badmintonracket}(F(x)) $;
            \item $ \neg \text{Likes}(x, \text{badminton})  \vee \text{Likes}(x, F(x))$;
            \item $ \neg \text{Badmintonracket}(x) \vee \neg \text{Likes(Alice}, x)$;
            \item $\text{Sport(badminton)}$;
            \item $\text{Sport(basketball)}$;
            \item $\neg \text{Sport}(x) \vee \neg \text{Likes}(y,x) \vee \neg \text{Likes}(z,x) \vee \text{Playtogether}(y,z) $;
            \item $ \neg \text{Playtogether}(x,y) \vee \text{Likes}(x, \text{badminton}) \vee \text{Likes}(x, \text{basketball})$;
            \item $ \neg \text{Playtogether}(x,y) \vee \text{Likes}(y, \text{badminton}) \vee \text{Likes}(x, \text{basketball})$;
            \item $ \neg \text{Playtogether}(x,y) \vee \text{Likes}(x, \text{badminton}) \vee \text{Likes}(y, \text{basketball})$;
            \item $ \neg \text{Playtogether}(x,y) \vee \text{Likes}(y, \text{badminton}) \vee \text{Likes}(y, \text{basketball})$;
            \item $ \text{Playtogether[Alice, Bob]}$;
            \item $ \neg \text{Playtogether[Father(Alice), Father(Bob)]}$;
            \item $ \text{Playtogether[Alice, Father(Alice)]}$;
            \item $ \text{Playtogether[Bob, Father(Bob)]}$;
            \item $ \neg \text{Likes(Bob, basketball)} \vee \neg \text{Likes(Bob, badminton)}$.

	\end{enumerate}

\begin{sidewaysfigure}[p]
        \centering
        \includegraphics[width=1\linewidth]{\figpath/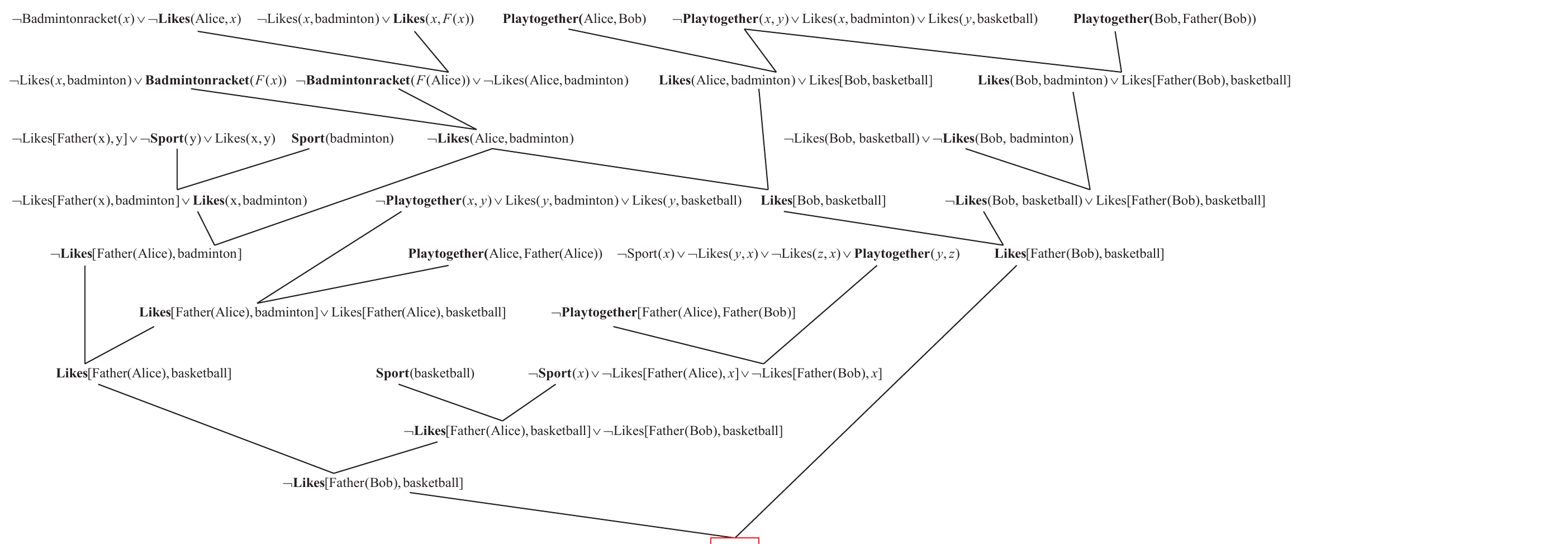}
		\caption{\label{fig-bob-like-basketball-badminton}The resolution proof showing that Bob likes basketball and badminton. At each resolution step, the unified literals are shown in bold.}
        \end{sidewaysfigure}

       For illustration, we give the detailed process of transforming the second sentence $ \forall x \, \text{Likes}(x, \text{badminton}) \to [\exists y \, \text{Badmintonracket}(y) \wedge \text{Likes}(x, y)] $ to CNF as an example:
        \begin{align}
		\nonumber  &\forall x \;\text{Likes}(x, \text{badminton}) \to [\exists y \; \text{Badmintonracket}(y) \wedge \text{Likes}(x, y)]  \Rightarrow \\
            \nonumber  &\forall x \exists y \; \neg \text{Likes}(x, \text{badminton}) \vee [ \text{Badmintonracket}(y) \wedge \text{Likes}(x, y)] \Rightarrow  \\
            \nonumber &\forall x \exists y \; [\neg \text{Likes}(x, \text{badminton}) \vee \text{Badmintonracket}(y)] \wedge [\neg \text{Likes}(x, \text{badminton})  \vee \text{Likes}(x, y)] \Rightarrow  \\
            \nonumber &\forall x \; [\neg \text{Likes}(x, \text{badminton}) \vee \text{Badmintonracket}(F(x))] \wedge [\neg \text{Likes}(x, \text{badminton})  \vee \text{Likes}(x, F(x))] \Rightarrow  \\
            &[\neg \text{Likes}(x, \text{badminton}) \vee \text{Badmintonracket}(F(x))] \wedge [\neg \text{Likes}(x, \text{badminton})  \vee \text{Likes}(x, F(x))].
	\end{align}
        The first transformation uses the relationship between implication and disjunction: $p \to q \equiv \neg p \vee q$. The second transformation is to distribute $\vee$ over $\wedge$ using Eq. (\ref{eq-distribute-disjunction}). The next transformation is the skolemization, where $F$ is a Skolem function depending on $x$. The last step is to remove the universal quantifier, since all remaining variables are implicitly universally quantified in CNF. The resolution proof that Bob’s father does not like basketball is given in Fig.~\ref{fig-bob-like-basketball-badminton}.

	Resolution in first-order logic has the additional step of unification compared to the resolution in propositional logic. This additional step, being computationally expensive on classical computers, poses a significant challenge to implementing quantum resolution in first-order logic. We address this problem by invoking the Herbrand’s theorem \cite{Russell2020Artificial}. This theorem states that: if a set $W$ of clauses is unsatisfiable, then there exists a finite subset of $H_W(W)$ that is also unsatisfiable. $H_W(W)$ refers to the Herbrand base of $W$. Given a set of clauses $W$,  then $H_W$, the Herbrand universe of $W$, is the set of all ground terms constructible from:
	\begin{enumerate}
		\item The function symbols in $W$;
		\item The constant symbols in $W$.
	\end{enumerate}
	Given a clause $ \neg P(x,F(x,A)) \vee \neg Q(x,A) \vee R(x,B) $, the corresponding Herbrand universe is:
	\begin{align}
		A,B,F(A,A),F(A,B),F(B,A),F(B,B), F(A,F(A,A)),\ldots
	\end{align}
	The Herbrand base of $W$ is the saturation of the set $W$ with respect to its Herbrand universe. The saturation is obtained by applying all possible substitutions of ground terms for variables in $W$. Thus, the Herbrand base of $ \neg P(x,F(x,A)) \vee \neg Q(x,A) \vee R(x,B) $ is:
	\begin{align}
	&\neg P(A,F(A,A))\vee \neg Q(A,A)\vee R(A,B), \\  \nonumber 
	&\neg P(B,F(B,A))\vee \neg Q(B,A)\vee R(B,B), \\ 	\nonumber 
	&\neg P(F(A,A),F(F(A,A),A))\vee \neg Q(F(A,A),A)\vee R(F(A,A),B), \\  \nonumber 
	&\neg P(F(A,B),F(F(A,B),A))\vee \neg Q(F(A,B),A)\vee R(F(A,B),B),\ldots
	\end{align}
    
    If a set of clauses $W$ in first-order logic is unsatisfiable, we use the iterative process of adopting increasingly larger subsets of $H_W(W)$. Since all the clauses have been propositionalized, we can use the same quantum resolution scheme described in Sec. \ref{section:quantum resolution in propositional logic} to perform quantum resolution of first-order logic on a quantum computer.
    In each iteration, we propositionalize the problem within these subsets and apply the quantum resolution introduced in Eq. (\ref{eq-quantum-resolution-qram}), continuing this process until unsatisfiability is demonstrated. If $W$ is satisfiable, no quantum or classical algorithm can decide the validity for all first-order logic sentences, a property known as the semi-decidability of first-order logic \cite{Russell2020Artificial}.

\section{Quantum automated theorem proving with Wu's method}

Compared with resolution-based reasoning, Wu’s method provides a more concrete and algebraic approach to automated theorem proving, and has proved highly effective for large classes of geometric problems~\cite{Chou198809introduction}. The central idea of Wu’s method is to transform a geometric statement into an equivalent algebraic problem: by introducing an appropriate coordinate system, all geometric relations appearing in the hypotheses and the conclusion are expressed as polynomial equations. The proof then proceeds by eliminating variables from the conclusion polynomial using the known hypothesis polynomials through a systematic process of pseudo-division. If all dependent variables can be eliminated and the conclusion reduces to the trivial identity $0=0$, the original geometric statement is proved.

A major computational challenge in Wu’s method arises from the rapid growth of intermediate multivariate polynomials. These polynomials may contain an exponential number of terms, leading to significant time and space overhead in classical implementations. Quantum computers, by contrast, can store and process information in superposition, suggesting a natural opportunity to accelerate algebraic reasoning. In this work, we propose a quantum approach to geometric theorem proving that implements the pseudo-division steps of Wu’s method directly on a quantum computer, thereby enabling the reasoning of geometric statements within a quantum framework.

\subsection{Preliminaries of Wu's method}

Before introducing the quantum implementation, we briefly review the essential idea of Wu’s method and its core algebraic ingredients. Wu’s method has been extensively developed and successfully applied in classical automated geometry proving~\cite{Chou198809introduction,Ye2008introduction}. Compared with quantum resolution, its algebraic structure makes it particularly amenable to near-term quantum realization, as it relies on arithmetic operations rather than deep logical branching.

The basic idea of Wu’s method is to prove elementary geometry statements by translating them into polynomial equations~\cite{Chou198809introduction}. An appropriate coordinate system is first chosen to parameterize all points in the geometric configuration. The independent coordinate variables are denoted by $u_1,\ldots,u_a$ (collectively written as $\mathbf{u}$), while the remaining variables $x_1,\ldots,x_b$ (collectively written as $\mathbf{x}$) are algebraically dependent on $\mathbf{u}$. Both the hypotheses and the conclusion of the geometric statement are then represented as polynomial equations in the variables $(\mathbf{u},\mathbf{x})$. Wu’s method provides a systematic procedure to derive the conclusion through successive pseudo-division of these polynomials.

In general, Wu’s method consists of three main steps. The first step converts the geometric statement into its algebraic form, yielding a set of polynomial equations. The second step performs \emph{triangulation} of the hypothesis polynomials via pseudo-division, so that each polynomial introduces exactly one new dependent variable. The final step sequentially pseudo-divides the conclusion polynomial by the triangulated hypotheses, eliminating dependent variables one by one. After all dependent variables have been eliminated, the resulting pseudo-remainder is either a polynomial in $\mathbf{u}$ or identically zero. In the latter case, the conclusion follows from the hypotheses under generic (non-degenerate) conditions. Such non-degeneracy assumptions correspond to exceptional parameter values that are geometrically insignificant and are standard in Wu’s method~\cite{Chou198809introduction}. As our focus is on the quantum implementation rather than geometric special cases, we restrict attention to the generic setting in which these conditions are satisfied.

The fundamental algebraic operation used throughout Wu’s method is \emph{pseudo-division}. Consider two multivariate polynomials $\text{S}$ and $\text{T}$, which are to be pseudo-divided with respect to the variable $y$, while the remaining variables are denoted by $x_1,x_2,\ldots,x_\text{K}$:
\begin{subequations}\label{eq-polynomials-st}
\begin{align}
    \text{S} &= \sum_{{\bf i},d_1}^{{\bf I},\text{D}_\text{s}}
    \text{s}({\bf i},d_1)\prod_{k=1}^\text{K} x_k^{i_k} y^{d_1}, \\
    \text{T} &= \sum_{{\bf j},d_2}^{{\bf J},\text{D}_\text{t}}
    \text{t}({\bf j},d_2)\prod_{k=1}^\text{K} x_k^{j_k} y^{d_2},
\end{align}
\end{subequations}
where $\text{s}(\cdot)$ and $\text{t}(\cdot)$ denote the coefficients of $\text{S}$ and $\text{T}$, respectively. The multi-indices ${\bf i}=(i_1,\ldots,i_\text{K})$ and ${\bf j}=(j_1,\ldots,j_\text{K})$ encode the degrees of the variables $x_k$. We denote by $\text{D}_\text{s}$ and $\text{D}_\text{t}$ the degrees of $\text{S}$ and $\text{T}$ in $y$, and by $\text{I}_k$ and $\text{J}_k$ the maximal degrees of the variables $x_k$.

Assuming $\text{D}_\text{s} > \text{D}_\text{t}$, the first step of pseudo-division computes
\begin{align}\label{eq-pdivsion}
    \text{R} = 
    \sum_{\bf j}^{\bf J} \text{t}({\bf j},\text{D}_\text{t})\prod_{k=1}^\text{K} x_k^{j_k} \text{S}
    -
    \sum_{\bf i}^{\bf I} \text{s}({\bf i},\text{D}_\text{s})\prod_{k=1}^\text{K} x_k^{i_k} \text{T}\, y^{\text{D}_\text{s}-\text{D}_\text{t}},
\end{align}
where the two prefactors $\sum_{\bf j}^{\bf J} \text{t}({\bf j},\text{D}_\text{t})\prod_{k=1}^\text{K} x_k^{j_k}$ and $\sum_{\bf i}^{\bf I} \text{s}({\bf i},\text{D}_\text{s})\prod_{k=1}^\text{K} x_k^{i_k}$ are the leading coefficients of $\text{T}$ and $\text{S}$ with respect to $y$, respectively.

Expanding Eq.~(\ref{eq-pdivsion}), the resulting polynomial $\text{R}$ can be written as
\begin{align}\label{eq-pdivsion-R}
    \text{R} = \sum_{{\bf i},{\bf j},d}^{{\bf I},{\bf J},\text{D}_\text{s}}
    \Bigl[
        \text{s}({\bf i},d)\,\text{t}({\bf j},\text{D}_\text{t})
        -
        \text{s}({\bf i},\text{D}_\text{s})\,\text{t}({\bf j},d+\text{D}_\text{t}-\text{D}_\text{s})
    \Bigr]
    \prod_{k=1}^\text{K} x_k^{i_k+j_k} y^d .
\end{align}
In particular, the coefficient of $y^{\text{D}_\text{s}}$ vanishes identically:
\begin{align}
    \sum_{{\bf i},{\bf j}}^{{\bf I},{\bf J}}
    \bigl[
        \text{s}({\bf i},\text{D}_\text{s})\text{t}({\bf j},\text{D}_\text{t})
        -
        \text{s}({\bf i},\text{D}_\text{s})\text{t}({\bf j},\text{D}_\text{t})
    \bigr]
    \prod_{k=1}^\text{K} x_k^{i_k+j_k}
    = 0,
\end{align}
which implies that $\deg(\text{R},y) < \max[\deg(\text{S},y),\deg(\text{T},y)]$. By iterating this procedure, the degree in $y$ is strictly reduced at each step, yielding the pseudo-remainder $\rm{prem}(\text{S},\text{T},y)$ satisfying
\begin{align}
    \deg [{\rm{prem}}({\rm{S}},{\rm{T}},y),y] < \min [\deg ({\rm{S}},y),\deg ({\rm{T}},y)].
\end{align}

This pseudo-division procedure is used to eliminate variables during the triangulation of the hypotheses and, subsequently, during the elimination of variables from the conclusion polynomial. In Wu’s method, the conclusion is proved once the final pseudo-remainder vanishes identically, completing the algebraic proof of the original geometric statement.

Now we analyze the complexity of Wu's method. From Eq.~(\ref{eq-pdivsion-R}), computing all coefficients of the polynomial $\text{R}$ in a single pseudo-division step requires, in the worst case,
\begin{align}
({{\rm{D}}_{\rm{s}}} + {{\rm{D}}_{\rm{t}}} + 2)\prod\limits_{k = 1}^{\rm{K}} {{\rm{(}}{{\rm{I}}_k}{\rm{ + 1)}}} \prod\limits_{k = 1}^{\rm{K}} {{\rm{(}}{{\rm{J}}_k} + 1{\rm{)}}} 
\end{align}
integer multiplications. Moreover, Eq.~(\ref{eq-pdivsion-R}) shows that the degree of $\text{R}$ is bounded above by 
\begin{align}
\text{D}_\text{s} - 1 + \sum_{k=1}^\text{K} (\text{I}_k + \text{J}_k),
\end{align}
which is essentially the sum of the degrees of the two input polynomials. Because the resulting polynomial $\text{R}$ is used as an input to subsequent pseudo-division steps, both the degree and the number of terms may increase rapidly as the procedure progresses, potentially leading to exponential growth.

A central metric for the complexity of the overall proof is therefore the maximum number of terms appearing among all intermediate polynomials generated during successive pseudo-divisions~\cite{Chou198809introduction}. For a polynomial of total degree $\text{D}$ in $\text{K}$ variables, the number of coefficients scales as the binomial coefficient $\binom{\text{D}+\text{K}}{\text{D}}$, which grows exponentially with both $\text{D}$ and $\text{K}$. Consequently, the worst-case time and space complexity of Wu’s method increases exponentially in the number of variables and the polynomial degree. Although Wu’s method is theoretically complete for equality-based geometric reasoning, this rapid growth in computational cost severely limits its practical applicability, motivating the exploration of alternative implementations—such as the quantum approach introduced in this work—that may better manage the combinatorial explosion inherent in algebraic geometry proving.

As a simple illustrative example, we apply Wu’s method to prove the classical geometric fact that the diagonals of a rhombus are perpendicular.

\begin{figure}[htb]
	\centering
	\includegraphics[width=0.25\linewidth]{\figpath/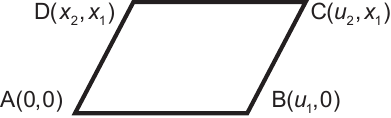}
	\caption{\label{fig-rhombus}A rhombus.}
\end{figure}

Let $\text{ABCD}$ be a rhombus, that is, a quadrilateral with both pairs of opposite sides parallel and all sides of equal length. As shown in Fig.~\ref{fig-rhombus}, we choose a coordinate system in which the four vertices have coordinates $\text{A}(0,0)$, $\text{B}(u_1,0)$, $\text{C}(u_2,x_1)$, and $\text{D}(x_2,x_1)$.
This choice reflects the assumptions that the sides $\text{AB}$ and $\text{CD}$ are parallel ($\text{AB}\parallel \text{CD}$) and of equal length ($\text{AB}=\text{CD}$). The remaining defining properties of a rhombus—namely $\text{AD}\parallel \text{BC}$ and $\text{AB}=\text{AD}$—can then be translated into the following polynomial equations:
\begin{subequations}\label{eq-rhombus-hypotheses}
\begin{align}
	h_1 &= -x_2 + (u_2 - u_1) = 0, \\
	h_2 &= -(x_1^2 + x_2^2) + u_1^2 = 0.
\end{align}
\end{subequations}

The geometric conclusion to be proved is that the diagonals $\text{AC}$ and $\text{BD}$ are perpendicular, denoted $text{AC} \perp \text{BD}$. In algebraic form, this condition can be expressed as
\begin{align}\label{eq-rhombus-conclusion}
	g_1 = x_1^2 + u_2(x_2 - u_1) = 0.
\end{align}
The problem is thus reduced to deriving the polynomial equation $g_1=0$ from the hypotheses $h_1=0$ and $h_2=0$ using algebraic manipulations. Wu’s method provides a systematic procedure for carrying out this derivation through successive pseudo-divisions.

Starting from the triangulated hypotheses in Eq.~(\ref{eq-rhombus-hypotheses}), the first pseudo-division step yields
\begin{align}\label{eq-rhombus-step1}
	r_1 = -g_1 - h_2
	= x_2^2 - u_2 x_2 + (u_1 u_2 - u_1^2).
\end{align}
The second pseudo-division eliminates the remaining quadratic term in $x_2$:
\begin{align}\label{eq-rhombus-step2}
	r_2 = -r_1 - x_2 h_1
	= u_1 x_2 + u_1 u_2 - u_1^2.
\end{align}
Finally, a third pseudo-division removes the remaining dependence on $x_2$:
\begin{align}\label{eq-rhombus-step3}
	r_3 = -r_2 - u_1 h_1 = 0.
\end{align}
The vanishing of the final pseudo-remainder $r_3$ completes the proof, establishing that the diagonals of any rhombus are perpendicular.

	\subsection{State-based quantum pseudo-division}

    In this section, we introduce a state-based approach for quantum pseudo-division of polynomials (QPP). In this setting, the QPP algorithm takes as input a quantum representation of two multivariate polynomials and outputs a quantum representation of their pseudo-remainder. In this subsection, the polynomials are represented by quantum states encoding their coefficients. The coefficients of the pseudo-remainder are extracted by measuring the output state in the Pauli-$Z$ basis. The measurement outcomes are then processed on a classical computer and re-encoded into quantum states for subsequent operations. Starting from two polynomials $\text{S}$ and $\text{T}$ defined in Eq.~(\ref{eq-polynomials-st}), the first step of the algorithm is to encode the coefficients of $\text{S}$ and $\text{T}$ into a single quantum state:
	\begin{align}\label{eq-state-st}
		\left| \psi  \right\rangle  = {1 \over c_0}\sum\limits_{{\bf{i}}}^{{\bf{I}}} \sum\limits_{{\bf{j}}}^{{\bf{J}}} \sum\limits_{d = 0}^{{\text{D}_\text{s}}}\left| d \right\rangle \left| {{\bf{i}}} \right\rangle \left| {\text{s}\left( {{\bf{i}},d} \right)} \right\rangle \left| {\text{s}\left( {{\bf{i}},{\text{D}_\text{s}}} \right)} \right\rangle \left| {{\bf{j}}} \right\rangle \left| {\text{t}\left( {{\bf{j}},d + {\text{D}_\text{t}} - {\text{D}_\text{s}}} \right)} \right\rangle  \left| {\text{t}\left( {{\bf{j}},{\text{D}_\text{t}}} \right)} \right\rangle .
	\end{align}
    Here $c_0$ is a normalization factor given by $c_0 = 2^{I_1 + \cdots + I_K + J_1 + \cdots + \text{D}_\text{s}}$.
    The indices ${\bf i}$ and ${\bf j}$ label the degrees of variables other than $y$, while $d$ denotes the degree of $y$.
    The registers $\left| \text{s}({\bf i},d) \right\rangle$ and $\left| \text{t}({\bf j},d+\text{D}_\text{t}-\text{D}_\text{s}) \right\rangle$ store the corresponding coefficients of $\text{S}$ and $\text{T}$, respectively.
    The registers $\left| \text{s}({\bf i},\text{D}_\text{s}) \right\rangle$ and $\left| \text{t}({\bf j},\text{D}_\text{t}) \right\rangle$ encode the leading coefficients of $\text{S}$ and $\text{T}$ with respect to the variable $y$, consistent with the classical pseudo-division procedure.
    The shifted index $d + \text{D}_\text{t} - \text{D}_\text{s}$ in the coefficients of $\text{T}$ reflects the standard pseudo-division step, where the polynomial of lower degree in $y$ is multiplied by $y^{\text{D}_\text{s}-\text{D}_\text{t}}$ prior to subtraction.
    This adjustment is implemented here as a coherent shift of the coefficient indices in the quantum state.

	In the quantum pseudo-division procedure, the first step is to multiply the polynomial $\text{S}$ by the leading coefficient of $\text{T}$ with respect to the variable $y$, and to multiply the polynomial $\text{T}$ by the leading coefficient of $\text{S}$ with respect to $y$. We introduce additional qubits to store the results of these multiplications. The result quantum state is given by:
	\begin{align}\label{eq-QPP-multiply}
		{1 \over {{c_0}}}  \sum\limits_{{\bf{i}}}^{{\bf{I}}} \sum\limits_{{\bf{j}}}^{{\bf{J}}} \sum\limits_{d = 0}^{{\text{D}_\text{s}}} \left| d \right\rangle \left| {{\bf{i}}} \right\rangle \left| {\text{s}\left( {{\bf{i}},d} \right)} \right\rangle \left| {\text{s}\left( {{\bf{i}},{\text{D}_\text{s}}} \right)} \right\rangle \left| {{\bf{j}}} \right\rangle \left| {\text{t}\left( {{\bf{j}},d + {\text{D}_\text{t}} - {\text{D}_\text{s}}} \right)} \right\rangle \left| {\text{t}\left( {{\bf{j}},{\text{D}_\text{t}}} \right)} \right\rangle
		\left| {\text{s}\left( {{\bf{i}},d} \right)\text{t}\left( {{\bf{j}},{\text{D}_\text{t}}} \right)} \right\rangle
		\left| {\text{t}\left( {{\bf{j}},d + {\text{D}_\text{t}} - {\text{D}_\text{s}}} \right)\text{s}\left( {{\bf{i}},{\text{D}_\text{s}}} \right)} \right\rangle .
	\end{align}
	For fixed indices ${\bf i}$, ${\bf j}$, and $d$, the last two registers encode the products $\text{s}({\bf i},d)\text{t}({\bf j},\text{D}_\text{t})$ and $\text{t}({\bf j},d+\text{D}_\text{t}-\text{D}_\text{s})\text{s}({\bf i},\text{D}_\text{s})$, respectively.
    The next step is to perform the subtraction between the two products and calculate the powers of different variables by adding the corresponding indices $\bf{i}+\bf{j}$. The result state is 
    \begin{align}\label{eq-QPP-sub} 
    \nonumber {1 \over {{c_0}}}\sum\limits_{\bf{i}}^{\bf{I}} {\sum\limits_{\bf{j}}^{\bf{J}} {\sum\limits_{d = 0}^{{\text{D}_\text{s}}} {\left| d \right\rangle } } } \left| {\bf{i}} \right\rangle & \left| {\text{s}\left( {{\bf{i}},d} \right)} \right\rangle \left| {\text{s}\left( {{\bf{i}},{\text{D}_\text{s}}} \right)} \right\rangle \left| {\bf{j}} \right\rangle \left| {\text{t}\left( {{\bf{j}},d + {\text{D}_\text{t}} - {\text{D}_\text{s}}} \right)} \right\rangle \left| {\text{t}\left( {{\bf{j}},{\text{D}_\text{t}}} \right)} \right\rangle \\ & \left| {\text{s}\left( {{\bf{i}},d} \right)\text{t}\left( {{\bf{j}},{\text{D}_\text{t}}} \right) - \text{t}\left( {{\bf{j}},d + {\text{D}_\text{t}} - {\text{D}_\text{s}}} \right)\text{s}\left( {{\bf{i}},{\text{D}_\text{s}}} \right)} \right\rangle \left| {\text{t}\left( {{\bf{j}},d + {\text{D}_\text{t}} - {\text{D}_\text{s}}} \right)\text{s}\left( {{\bf{i}},{\text{D}_\text{s}}} \right)} \right\rangle \left| {{\bf{i}} + {\bf{j}}} \right\rangle . 
    \end{align}
    
    A classical description of the pseudo-remainder of $\text{S}$ divided by $\text{T}$ with respect to the variable $y$ is obtained by measuring the resulting state in the Pauli-$Z$ basis. The measurement outcomes of the registers $\left|{\bf i}+{\bf j}\right\rangle$ and $\left|d\right\rangle$ specify the powers of the variables, while the outcomes of the coefficient register yield the corresponding coefficients.
    However, for a fixed multi-index ${\bf i}+{\bf j}$, the associated coefficients are stored in a superposition over different decompositions of ${\bf i}$ and ${\bf j}$. Recovering the final coefficient therefore requires summing all contributions corresponding to the same monomial on a classical computer, which is equivalent to combining like terms in classical polynomial multiplication. Specifically, the state corresponding to ${\bf i}+{\bf j}={\bf I}_0$ is:
    \begin{align}
    \label{eq-coe-i0}
    \sum\limits_{\bf{i}}^{{\bf{I}}_0} {\left| {\bf{i}} \right\rangle } \left| {\text{s}\left( {{\bf{i}},d} \right)t\left( {{{\bf{I}}_0} - {\bf{i}},{\text{D}_\text{t}}} \right) - t\left( {{{\bf{I}}_0} - {\bf{i}},d + {\text{D}_\text{t}} - {\text{D}_\text{s}}} \right)\text{s}\left( {{\bf{i}},{\text{D}_\text{s}}} \right)} \right\rangle, 
    \end{align}
    where the normalization factor is omitted. The coefficient corresponding to $({\bf I}_0,d)$ is obtained by measuring all terms in Eq.~(\ref{eq-coe-i0}) and summing them over different values of ${\bf i}$.
    As a consequence, the state-based coefficient encoding requires measurement and classical post-processing to combine like terms after each polynomial multiplication. This measurement-induced collapse prevents coherent reuse of intermediate results and therefore would hardly offer any quantum advantage over classical pseudo-division. In this work, we will not use this state-based approach, but  instead exploit a more convenient circuit-based approach. The state-based approach introduced here serves merely as a conceptual baseline for comparison.

\subsection{Circuit-based quantum pseudo-division}

We now introduce a circuit-based framework for quantum pseudo-division, designed to perform all algebraic operations required by Wu’s method using only a single encoding step at the beginning of the computation and a single measurement step at the end. In contrast to the state-based approach discussed above, polynomials in this framework are encoded directly into quantum circuits and manipulated coherently throughout the entire proof procedure. The construction is based on the point–value representation of polynomials and therefore avoids the explicit aggregation of like terms on classical hardware, which constitutes a fundamental bottleneck for state-based implementations.

This subsection is organized as follows. We first describe how multivariate polynomials can be encoded as quantum circuits. We then explain how polynomial interpolation can be realized within the circuit model to extract coefficient polynomials with respect to a chosen variable. Finally, we show how these components are combined to implement quantum pseudo-division and to support the iterative variable-elimination steps required by Wu’s method.

\subsubsection{Encoding polynomials into quantum circuits}

We begin by describing how a multivariate polynomial can be represented by a quantum circuit. Taking the polynomial $\text{S}$ in Eq.~(\ref{eq-polynomials-st}) as an example, we represent $\text{S}$ by a quantum circuit $U_\text{S}$. The circuit takes the variables $\mathbf{x}$ and $y$ as inputs, each encoded as a binary string on quantum registers, together with an auxiliary register initialized in the state $\lvert \mathbf{0} \rangle$. Here, $\mathbf{x}$ denotes the collection of variables $\{x_k\}$, and $\lvert \mathbf{0} \rangle$ is a shorthand for the multi-qubit all-zero state $\lvert 00\cdots 0 \rangle$. The circuit evaluates the polynomial $\text{S}$ and writes the result into the auxiliary register as a quantum state encoding the corresponding binary integer:
\begin{align}\label{eq-us}
    U_\text{S} \lvert \mathbf{x} \rangle \lvert y \rangle \lvert \mathbf{0} \rangle
    = \lvert \mathbf{x} \rangle \lvert y \rangle \lvert \text{S}(\mathbf{x},y) \rangle .
\end{align}

The circuit-based representation in Eq.~(\ref{eq-us}) is naturally compatible with the point–value representation of polynomials. In classical computation, the point–value representation is a standard technique for accelerating polynomial multiplication, reducing the time complexity from $\Theta(\text{D}^2)$ to $\Theta(\text{D}\log \text{D})$ for degree-$\text{D}$ univariate polynomials. In this representation, two degree-$\text{D}$ polynomials $f$ and $h$ are specified by $\text{D}+1$ point–value pairs $(x_d,f(x_d))$ and $(x_d,h(x_d))$. Their product $f\cdot h$, which has degree at most $2\text{D}$, is uniquely determined by $2\text{D}{+}1$ point–value pairs $(x_d,f(x_d)h(x_d))$. Polynomial multiplication is thereby reduced to a collection of independent pointwise multiplications, without ever forming or aggregating monomial coefficients.

This feature is crucial for quantum computation. Combining like terms is an irreversible operation and therefore fundamentally incompatible with coherent quantum evolution. By working entirely within the point–value representation and constructing new circuit representations directly from existing ones, repeated pseudo-division steps can be carried out coherently, without intermediate measurements, classical post-processing, or re-encoding of polynomials.

We adopt this perspective to encode polynomials into quantum circuits and consider two natural strategies for realizing a point–value circuit representation on a quantum device. The first strategy is \emph{arithmetic-based}: the circuit explicitly implements the algebraic structure of the polynomial using quantum integer addition, subtraction, and multiplication, enabling evaluation on arbitrary integer inputs encoded as qubit strings. This approach follows directly from the symbolic expression of the polynomial and has been illustrated with concrete examples in the main text. The second strategy is \emph{data-driven}: a sufficient set of input points is selected in advance, the corresponding polynomial values are classically computed, and these index–value pairs are embedded directly into a quantum circuit. The circuit then maps an index register labeling the sampled points to a register storing the associated polynomial values. 
Each strategy has distinct advantages. The arithmetic-based approach supports evaluation on arbitrary inputs and aligns closely with symbolic algebra, whereas the data-driven approach offers greater flexibility in the choice of sampling points and often leads to simpler circuit constructions.

We now briefly analyze the circuit complexity associated with these two encoding strategies. For the arithmetic-based approach, the circuit cost is determined by the total cost of evaluating and accumulating all nonzero monomials in the polynomial. Each monomial is evaluated independently from qubit-string–encoded inputs using quantum integer multiplication, together with the necessary additions and shifts. Consequently, the overall gate count and circuit depth are obtained by summing the costs of the arithmetic subcircuits corresponding to each monomial, rather than by a single uniform per-term prefactor. The dominant cost arises from quantum integer multiplication. The best known ancilla-free constructions achieve an asymptotic gate complexity of $\mathcal{O}(n^{1+\epsilon})$ for multiplying $n$-bit integers, for any $\epsilon>0$, with practical constructions exhibiting scalings as low as $\mathcal{O}(n^{1.3})$.

For the data-driven strategy, the dominant cost arises from loading classical point–value data into a quantum circuit. Once the point values are computed classically, the quantum circuit needs only to implement a data-loading unitary that maps an index register to the corresponding stored value. Using a construction analogous to the $U_\text{KB}$ circuit introduced in the main text, such a unitary can be implemented with circuit depth scaling as $O(\text{D}\log \text{D})$, where $\text{D}$ is the number of point–value pairs used to represent the polynomial. Subsequent polynomial operations, including multiplication and repeated pseudo-division steps, can then be performed entirely within the point–value representation, without introducing additional overhead associated with combining like terms.

\subsubsection{Quantum polynomial interpolation}

To eliminate a specific variable $y$ in the QPP procedure, we introduce a quantum circuit $U_{\text{S},y}$ that outputs the coefficient of $y^d$ in the polynomial $\text{S}$, with the exponent $d$ provided as an input. The construction of $U_{\text{S},y}$ from the evaluation circuit $U_\text{S}$ is achieved via a quantum polynomial interpolation procedure, which we describe in detail below. The circuit implements the following transformation:
\begin{align}\label{eq-usy}
    U_{\text{S},y}\lvert d \rangle \lvert \mathbf{x} \rangle \lvert \mathbf{0} \rangle
    =
    \lvert d \rangle \lvert \mathbf{x} \rangle
    \left|
        \sum_{\bf i}^{\bf I} \text{s}({\bf i},d)\prod_{k=1}^\text{K} x_k^{i_k}
    \right\rangle .
\end{align}
Here, $\sum_{\bf i}^{\bf I} \text{s}({\bf i},d)\prod_k^\text{K} x_k^{i_k}$ is precisely the coefficient of $y^d$ in $\text{S}$ as defined in Eq.~(\ref{eq-polynomials-st}). Importantly, this coefficient is itself a multivariate polynomial in the remaining variables $\mathbf{x}$.

As a simple illustration, the coefficient of $y^2$ ($d=2$) in the polynomial $x_1^2 y^2 + x_2 y^2 - y^2 + x_1 y$ is the polynomial $x_1^2 + x_2 - 1$. A schematic representation of the circuit $U_{\text{S},y}$ is shown in Fig.~\ref{fig-qp-unitary-detail}\textbf{a}. For clarity, the registers encoding the variables $\mathbf{x}$ are omitted, and the output state $\left| \sum_{\bf i}^{\bf I} \text{s}({\bf i},d)\prod_k\text{K} x_k^{i_k} \right\rangle$ is abbreviated as $\lvert \text{s}({\bf i},d)\rangle$.

\paragraph{Interpolation strategy.}
We now describe how to construct the circuit $U_{\text{S},y}$ from the evaluation circuit $U_\text{S}$. This task can be viewed as a univariate polynomial interpolation problem in the variable $y$, whose goal is to extract the coefficient polynomials associated with different powers of $y$ in $\text{S}(\mathbf{x},y)$. If $\text{S}(\mathbf{x},y)$ has degree $\text{D}$ in $y$, then determining $U_{\text{S},y}$ amounts to recovering $D{+}1$ coefficient polynomials in the remaining variables $\mathbf{x}$.

Classically, this problem is typically solved by evaluating the polynomial at a suitable set of interpolation points $\{y_d\}$ and reconstructing the coefficients from these values. A common choice is Fourier-based interpolation, in which the points are chosen as the $\text{D}$-th roots of unity $y_d = \exp(2\pi i d/\text{D})$, enabling the use of fast Fourier transform algorithms. In the quantum setting, however, such approaches require complex-valued arithmetic and floating-point representations, which introduce substantial overhead.

\paragraph{Kravchuk-based quantum interpolation.}
To avoid these complications, we adopt the \emph{Kravchuk polynomial basis}~\cite{Szegoe1939Orthogonal} to perform interpolation entirely over integer-valued domains. This choice preserves real-valued arithmetic throughout the computation and is therefore directly compatible with our qubit-string encoding of integers, eliminating the need for complex-number manipulation on quantum hardware~\cite{Seidel2022Efficient}.

The Kravchuk polynomials are defined as
\begin{align}\label{eq-kravchuk-poly}
    K_d(y,\text{D},q)
    =
    \sum_{j=0}^d (-1)^j (q-1)^{d-j}
    C(\text{D}-j,d-j) C(y,j),
\end{align}
where $C(\cdot,\cdot)$ denotes the binomial coefficient. In this work, we restrict ourselves to the simplest and most practical case $q=2$. The Kravchuk polynomials satisfy the orthogonality relation:
\begin{align}\label{eq-kravchuk-orth}
    \sum_{d=0}^\text{D}
    C(\text{D},d)\,K_e(d,\text{D},2)\,K_f(d,\text{D},2)
    =
    2^\text{D} C(\text{D},e)\,\delta_{e,f}.
\end{align}
From this orthogonality, the corresponding completeness relation follows:
\begin{align}
    \sum\limits_{e = 0}^{\rm{D}} {{1 \over {C({\rm{D}},e)}}{K_e}(d,{\rm{D}},2){\mkern 1mu} {K_e}(f,{\rm{D}},2)}  = {{{2^{\rm{D}}}} \over {C({\rm{D}},d){\mkern 1mu} }}{\delta _{d,f}}.
\end{align}

Given a univariate polynomial $\text{S}(y)$ evaluated at integer points $y=d$, where the dependence on $\mathbf{x}$ is suppressed for clarity, the corresponding Kravchuk coefficients are given by
\begin{align}\label{eq-kravchuk-cal-coe}
    c_e
    =
    \sum_{d=0}^\text{D}
    C(\text{D},d)\,\text{S}(d)\,K_e(d,\text{D},2).
\end{align}
The polynomial $\text{S}(y)$ can then be reconstructed via the expansion
\begin{align}\label{eq-kravchuk-decompose-coe}
    \text{S}(y)
    =
    \sum_{d=0}^\text{D}
    \frac{c_d}{2^\text{D} C(\text{D},d)}\,K_d(y,\text{D},2).
\end{align}

\paragraph{Quantum implementation.}
In the quantum setting, computing the Kravchuk coefficients $\lvert c_d\rangle$—rather than the monomial-basis coefficients $\lvert \text{s}({\bf i},d)\rangle$—can be decomposed into three coherent components:
\begin{enumerate}
    \item preparing the integer weights $\lvert C(\text{D},y)K_d(y,\text{D},2)\rangle$,
    \item generating the point values $\lvert \text{S}(0)\rangle,\lvert \text{S}(1)\rangle,\ldots,\lvert \text{S}(\text{D})\rangle$ using the evaluation circuit $U_\text{S}$, and
    \item applying a reversible arithmetic circuit to compute the weighted sum
    $\lvert \sum_{y=0}^\text{D} C(\text{D},y)\text{S}(y)K_d(y,\text{D},2)\rangle$.
\end{enumerate}

Since the factors $C(\text{D},y)K_d(y,\text{D},2)$ are independent of the polynomial values $\text{S}(y)$, they can be encoded using a fixed circuit $U_{CK}$ that prepares the corresponding integer data. This circuit is analogous to the data-loading procedure in Eq.~(\ref{eq-qram-u}) and has depth $O(\text{D}\log \text{D})$. Together with the evaluation circuit $U_\text{S}$, this enables the coherent computation of the Kravchuk coefficients.
    Specifically, the circuit allocates $\text{D}{+}1$ value registers, each associated with one evaluation point $y\in{0,1,\ldots,\text{D}}$, and uses $U_\text{S}$ to populate these registers with the point values $\text{S}(0),S(1),\ldots,S(\text{D})$. It then applies a reversible interpolation transform that computes the Kravchuk coefficient indexed by $d$ via the weighted sum over the populated value registers. This transform serves as the quantum analogue of classical coefficient extraction and yields the coefficient polynomial in encoded form. As a result, the circuit $U_{\text{S},y}$ outputs the coefficient polynomial corresponding to the specified index $d$ in a single coherent procedure, without intermediate measurements.

    \subsubsection{Quantum pseudo-division by constructing remainder circuit}

    \begin{figure}[htb]
		\includegraphics[width=1\linewidth]{\figpath/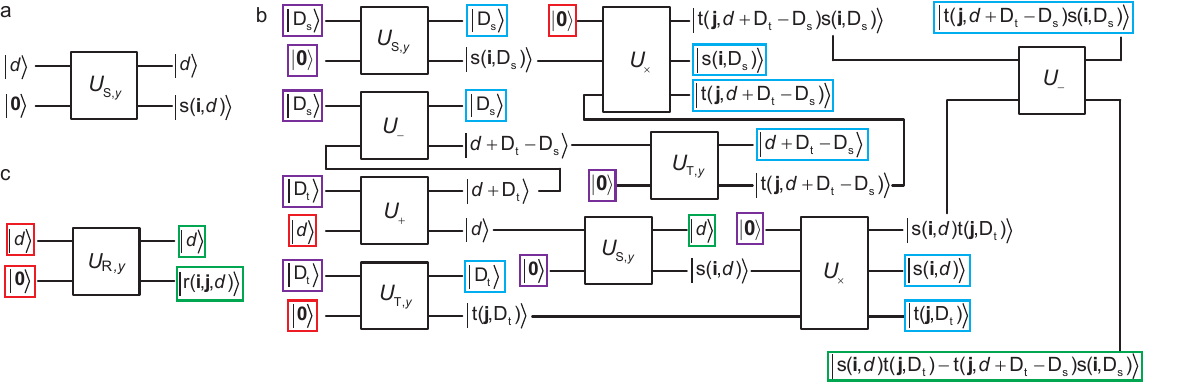}
		\caption{\label{fig-qp-unitary-detail}The quantum circuit $U_{\text{R},y}$ representing the polynomial $\text{R}$ obtained by pseudo-dividing $\text{S}$ by $\text{T}$ with respect to the variable $y$, as defined in Eq.~(\ref{eq-pdivsion-R}).  \textbf{a}, The quantum circuit of ${U_{\text{S},y}}$ in Eq. (\ref{eq-usy}).  The qubits encoding the variables $\bf{x}$ are omitted for brevity, and the state $\left| {\sum\limits_{\bf{i}}^{\bf{I}} {\text{s}({\bf{i}},d)\prod\limits_k^\text{K} {x_k^{{i_k}}} } } \right\rangle $ is abbreviated as $\left| {\text{s}({\bf{i}},d)} \right\rangle $. \textbf{b}, Constructing ${U_{\text{R},y}}$ with ${U_{\text{S},y}}$, ${U_{\text{T},y}}$, and basic arithmetic circuits. The circuits $U_{+}$, $U_{-}$, and $U_{\times}$ implement addition, subtraction, and multiplication for the input states, respectively. The states in the red boxes are the input states and those in the green boxes are the output states of ${U_{\text{R},y}}$. The remaining qubits are auxiliary qubits that are prepared onto the desired states (purple boxes) with one layer of single-qubit gates. These auxiliary qubits that end with states in blue boxes can be reset to $\left| {\bf{0}} \right\rangle $ by applying appropriate inverse circuits. \textbf{c}, Simplified representation of the quantum circuit $U_{\text{R},y}$, serving as a compact abbreviation of the full construction shown in \textbf{b}.}
    \end{figure}

        We use the two polynomials $\text{S}$ and $\text{T}$ defined in Eq.~(\ref{eq-polynomials-st}) to illustrate the circuit-based quantum pseudo-division procedure. The polynomial $\text{R}$, obtained by pseudo-dividing $\text{S}$ by $\text{T}$ with respect to the variable $y$, is represented by the quantum circuit $U_{\text{R},y}$. As shown in Fig.~\ref{fig-qp-unitary-detail}\textbf{b}, this circuit is assembled from $U_{\text{S},y}$ and $U_{\text{T},y}$ together with a set of basic arithmetic circuits. In Fig.~\ref{fig-qp-unitary-detail}\textbf{b}, registers shown in red correspond to input states, while registers shown in green correspond to the output states of $U_{\text{R},y}$. All remaining registers are initialized to predetermined auxiliary states, indicated by the purple boxes, which can be prepared using a single layer of single-qubit gates. For example, when $\text{D}_\text{t}=5$, the state $\lvert \text{D}_\text{t}\rangle=\lvert 5\rangle=\lvert 101\rangle$ can be prepared from $\lvert 000\rangle$ using two Pauli-$X$ gates. Executing the circuit yields the coefficient of $y^d$ in $\text{R}$, namely $\sum\limits_{{\bf{i}},{\bf{j}}}^{{\bf{I}},{\bf{J}}} {\text{r}({\bf{i}},{\bf{j}},d)\prod\limits_k^\text{K} {x_k^{{i_k} + {j_k}}} }  = \sum\limits_{{\bf{i}},{\bf{j}}}^{{\bf{I}},{\bf{J}}} {\left[ {\text{s}({\bf{i}},d)\text{t}({\bf{j}},{\text{D}_\text{t}}) - \text{t}({\bf{j}},d + {\text{D}_\text{t}} - {\text{D}_\text{s}})\text{s}({\bf{i}},{\text{D}_\text{s}})} \right]\prod\limits_k^\text{K} {x_k^{{i_k} + {j_k}}} } $ shown in the green box.

        \begin{figure}[htb]
		\includegraphics[width=0.7\linewidth]{\figpath/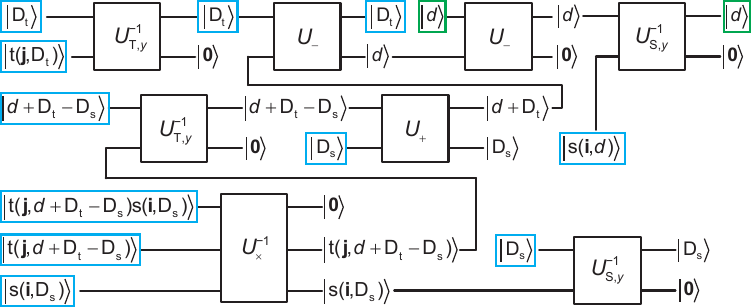}
		\caption{\label{fig-reset-auxiliary-ury}The quantum circuit that resets the auxiliary qubits in the construction of $U_{\text{R},y}$ shown in Fig.~\ref{fig-qp-unitary-detail}\textbf{b}. The states in the blue boxes directly correspond to the terminal state of auxiliary qubits in Fig.~\ref{fig-qp-unitary-detail}\textbf{b}. After the reset procedure, each multi-qubit auxiliary register represented by a single wire in this figure is in one of the states $\left| \bf{0} \right\rangle $, $\left| {{\text{D}_\text{s}}} \right\rangle $, and $\left| {{\text{D}_\text{t}}} \right\rangle $. All of these states can be reset to $\left| \mathbf{0} \right\rangle$ using no more than one layer of single-qubit gates. The state $\left| d \right\rangle $ in the green boxes is the output state of $U_{\text{R},y}$, which remain unchanged in this reset process.}
        \end{figure}

        At this stage, the auxiliary qubits may be reset for subsequent operations, although such a reset is not strictly necessary. After the circuit in Fig.~\ref{fig-qp-unitary-detail}\textbf{b} is executed, the auxiliary qubits are in the states $\left| {{\text{D}_\text{s}}} \right\rangle $, $\left| {{\text{D}_\text{t}}} \right\rangle $, $\left| {d + {\text{D}_\text{t}} - {\text{D}_\text{s}}} \right\rangle $, $\left| {\text{s}\left( {{\bf{i}},{\text{D}_\text{s}}} \right)} \right\rangle $, $\left| {\text{t}\left( {{\bf{j}},{\text{D}_\text{t}}} \right)} \right\rangle $, $\left| {\text{s}\left( {{\bf{i}},d} \right)} \right\rangle $, $\left| {\text{t}\left( {{\bf{j}},d + {\text{D}_\text{t}} - {\text{D}_\text{s}}} \right)} \right\rangle $, and $\left| {\text{t}\left( {{\bf{j}},d + {\text{D}_\text{t}} - {\text{D}_\text{s}}} \right)\text{s}\left( {{\bf{i}},{\text{D}_\text{s}}} \right)} \right\rangle $, as shown in the blue boxes in Fig.~\ref{fig-qp-unitary-detail}\textbf{b}. These qubits can be reset to $\lvert \bf{0}\rangle$ using the unitaries already present in Fig.~\ref{fig-qp-unitary-detail}\textbf{b} and their inverses. The explicit reset procedure is shown in Fig.~\ref{fig-reset-auxiliary-ury}.

\section{Polynomial identity testing on quantum computers}

Polynomial identity testing (PIT) is the problem of deciding whether a multivariate polynomial $\text{P}(x_1,\ldots,x_n)$ over a field $\mathbf{F}$ is identically zero. In computational settings, $\text{P}$ is typically specified implicitly, for example via an arithmetic circuit or a straight-line program, and the algorithm is allowed to evaluate $\text{P}$ at chosen input points. The goal of PIT is to determine whether $\text{P}\equiv 0$ using as few evaluations as possible \cite{Demillo197806probabilistic,Schwartz198010Fast,Zippel1979Probabilistic}. The algorithmic study of polynomial identity testing naturally splits into two distinct categories: randomized PIT and deterministic PIT. In the randomized setting, the algorithm is allowed to use randomness to select evaluation points and may err with small probability. This model leads to simple and efficient algorithms based on random sampling and underlies most practical PIT procedures. In contrast, deterministic PIT asks for a fixed, explicitly constructed set of evaluation points that certifies whether a given polynomial is identically zero, with no probability of error. Despite its conceptual simplicity, deterministic PIT is significantly more challenging and is closely connected to open problems in algebraic complexity theory \cite{Kabanets2003Derandomizing}.

The most widely used randomized algorithms for PIT are based on the Schwartz--Zippel lemma \cite{Schwartz198010Fast,Zippel1979Probabilistic,Demillo197806probabilistic} and its variants, which guarantee correctness with high probability using only a small number of random evaluations. These algorithms run in polynomial time and achieve exponentially small error probabilities with minimal randomness. On the other hand, no general polynomial-time deterministic algorithm is known for PIT when the polynomial is given by an arithmetic circuit.
Strong derandomization of PIT is known to imply circuit lower bounds \cite{Kabanets2003Derandomizing}. In particular, a deterministic polynomial-time PIT algorithm for general arithmetic circuits would imply that either $NEXP \not\subset P/poly$ or the permanent does not admit polynomial-size arithmetic circuits.

The central tool underlying randomized polynomial identity testing is the Schwartz--Zippel lemma. Let $\text{P}(x_1,\ldots,x_n)$ be a nonzero multivariate polynomial over a field $\mathbf{F}$ with total degree at most $\text{D}$, and let $\text{G}\subseteq\mathbf{F}$ be a finite subset. The Schwartz--Zippel lemma states that if a point $\mathbf{a}=(a_1,\ldots,a_n)$ is sampled uniformly at random from $\text{G}$, then the probability that $\text{P}(\mathbf{a})=0$ is at most $\text{D}/|\text{G}|$. This bound depends only on the total degree of the polynomial and the size of the sampling set, and is independent of the number of variables.

The proof of the Schwartz--Zippel lemma proceeds by induction on the number of variables $n$. In the base case $n=1$, the statement reduces to the elementary fact that a nonzero univariate polynomial of degree at most $\text{D}$ has at most $\text{D}$ roots. Consequently, when sampling uniformly from $\text{G}$, the probability of evaluating to zero is at most $\text{D}/|\text{G}|$.
For the inductive step, we consider a nonzero polynomial $\text{P}(x_1,\ldots,x_n)$ of total degree at most $\text{D}$, and regard it as a univariate polynomial in the variable $x_n$. We write $\text{P}(x_1,\ldots,x_n)=\sum_{k=0}^{d} \text{Q}_k(x_1,\ldots,x_{n-1}) x_n^k$, where $d\le \text{D}$ is the degree of $\text{P}$ in $x_n$, and the leading coefficient $\text{Q}_d$ is not identically zero. Each coefficient polynomial $\text{Q}_k$ depends only on the first $n-1$ variables.
Now sample $(a_1,\ldots,a_{n-1})$ uniformly at random from $\text{G}$. By the induction hypothesis, the probability that $\text{Q}_d(a_1,\ldots,a_{n-1})=0$ is at most $(\text{D}-d)/|\text{G}|$, since the total degree of $\text{Q}_d$ is at most $\text{D}-d$. Conditioned on the event that $\text{Q}_d(a_1,\ldots,a_{n-1})\neq 0$, the polynomial $\text{P}(a_1,\ldots,a_{n-1},x_n)$ becomes a nonzero univariate polynomial in $x_n$ of degree $d$, which therefore has at most $d$ roots. Sampling $a_n$ uniformly from $\text{G}$ then yields $\text{P}(a_1,\ldots,a_n)=0$ with probability at most $d/|\text{G}|$. Combining these two cases gives an overall failure probability bounded by $\text{D}/|\text{G}|$, completing the inductive proof.

The Schwartz--Zippel lemma immediately yields an efficient randomized algorithm for PIT. One samples a point uniformly from $\text{G}$ and evaluates $\text{P}$ at that point. If the result is nonzero, one can conclude that $\text{P}$ is not identically zero. Repeating this procedure independently $m$ times reduces the error probability to at most $(\text{D}/|\text{G}|)^m$. Thus, by choosing $|\text{G}|$ sufficiently large and repeating the test a logarithmic number of times, one obtains an algorithm with exponentially small error probability using only a small number of evaluations.

Despite the effectiveness of randomized algorithms, deterministic polynomial identity testing remains a fundamental open problem. In the deterministic setting, one must specify a fixed evaluation set that certifies nonzeroness for all nonzero polynomials in a given circuit class, equivalently an explicit hitting set. For general arithmetic circuits, no polynomial-time deterministic PIT algorithm is known. Moreover, deterministic PIT for general arithmetic circuits is widely believed to be computationally intractable, even in the presence of nondeterminism.
Deterministic PIT has been achieved only for restricted circuit classes, such as sparse polynomials and depth-bounded circuits, where additional algebraic structures can be exploited to construct explicit hitting sets \cite{Shpilka201012Arithmetic}. These results do not extend to general arithmetic circuits and highlight the sharp contrast between randomized and deterministic regimes.

Polynomial identity testing on quantum computers can be formulated in a manner analogous to the classical randomized approach but with the evaluation procedure embedded coherently into a quantum circuit. Given a quantum circuit $U_\text{P}$ that evaluates a polynomial $\text{P}$ on inputs from a finite set $\text{G}$, the task of PIT reduces to deciding whether there exists a point $\mathbf{a}\in G$ such that $\text{P}(\mathbf{a})\ne 0$.
Although quantum algorithms can offer significant speedups for certain search and sampling tasks, they do not fundamentally alter the derandomization landscape of polynomial identity testing. The core obstacle in deterministic PIT lies in the explicit construction of hitting sets that certify nonzeroness for all nonzero polynomials in a given circuit class. This requirement is inherently combinatorial and algebraic, rather than algorithmic, and is not addressed by quantum algorithms \cite{Montanaro201601Quantum}.

In the randomized setting, PIT already admits highly efficient classical algorithms based on the Schwartz–Zippel lemma, achieving exponentially small error probabilities with only a constant number of evaluations. In this regime, quantum algorithms cannot provide an asymptotic improvement, since the classical query complexity is already minimal. On the other hand, in the deterministic setting, quantum computation does not circumvent the need for explicit hitting sets. Quantum search algorithms, such as the Grover algorithm, can only provide probabilistic guarantees and therefore cannot yield deterministic identity tests.
Consequently, while quantum algorithms can yield a quadratic improvement in query complexity when searching for nonzero evaluations over a prescribed finite set, they do not resolve the central challenge of deterministic PIT. In particular, quantum computation does not bypass the structural barriers underlying derandomization, nor does it eliminate the complexity-theoretic consequences associated with deterministic polynomial identity testing.

\end{document}